\documentclass[iop]{emulateapj}
\usepackage{graphicx}
\usepackage{amsmath}
\usepackage{amssymb}
\usepackage{epstopdf}
\usepackage{natbib}
\usepackage{color}
\usepackage{hhline}
\usepackage{subfigure}
\usepackage{xspace}
\usepackage{hyperref}
\hypersetup{
     colorlinks   = true,
     allcolors    = blue
}

\usepackage{mymacros}
\usepackage{affils} 

\begin{document}

\title{First Season MWA EoR Power Spectrum Results at Redshift 7}
\shortauthors{Beardsley et al.}
\shorttitle{First Season MWA EoR}

\def\myemail{\altaffilmark{*}}
\def\myemailtxt{\altaffiltext{*}{e-mail: Adam.Beardsley@asu.edu}}

\DeclareAffil{ASU}{Arizona State University, School of Earth and Space Exploration, Tempe, AZ 85287, USA}
\DeclareAffil{email}{e-mail: Adam.Beardsley@asu.edu}
\DeclareAffil{UW}{University of Washington, Department of Physics, Seattle, WA 98195, USA}
\DeclareAffil{UWeSci}{University of Washington, eScience Institute, Seattle, WA 98195, USA}
\DeclareAffil{Brown}{Brown University, Department of Physics, Providence, RI 02912, USA}
\DeclareAffil{SKASA}{Square Kilometre Array South Africa (SKA SA), Park Road, Pinelands 7405, South Africa}
\DeclareAffil{RU}{Department of Physics and Electronics, Rhodes University, Grahamstown 6140, South Africa}
\DeclareAffil{CfA}{Harvard-Smithsonian Center for Astrophysics, Cambridge, MA 02138, USA}
\DeclareAffil{ANU}{Australian National University, Research School of Astronomy and Astrophysics, Canberra, ACT 2611, Australia}
\DeclareAffil{CAASTRO}{ARC Centre of Excellence for All-sky Astrophysics (CAASTRO)}
\DeclareAffil{Haystack}{MIT Haystack Observatory, Westford, MA 01886, USA}
\DeclareAffil{MIT}{MIT Kavli Institute for Astrophysics and Space Research, Cambridge, MA 02139, USA}
\DeclareAffil{Curtin}{International Centre for Radio Astronomy Research, Curtin University, Perth, WA 6845, Australia}
\DeclareAffil{USydney}{The University of Sydney, Sydney Institute for Astronomy, School of Physics, NSW 2006, Australia}
\DeclareAffil{Dunlap}{Dunlap Institute for Astronomy and Astrophysics, University of Toronto, ON M5S 3H4, Canada}
\DeclareAffil{Victoria}{Victoria University of Wellington, School of Chemical \& Physical Sciences, Wellington 6140, New Zealand}
\DeclareAffil{UWisc}{University of Wisconsin--Milwaukee, Department of Physics, Milwaukee, WI 53201, USA}
\DeclareAffil{UMichigan}{University of Michigan, Department of Atmospheric, Oceanic and Space Sciences, Ann Arbor, MI 48109, USA}
\DeclareAffil{UMelbourne}{The University of Melbourne, School of Physics, Parkville, VIC 3010, Australia}
\DeclareAffil{CASS}{CSIRO Astronomy and Space Science (CASS), PO Box 76, Epping, NSW 1710, Australia}
\DeclareAffil{Tata}{National Centre for Radio Astrophysics, Tata Institute for Fundamental Research, Pune 411007, India}
\DeclareAffil{ASTRON}{Netherlands Institute for Radio Astronomy (ASTRON), PO Box 2, 7990 AA Dwingeloo, The Netherlands}
\DeclareAffil{RRI}{Raman Research Institute, Bangalore 560080, India}
\DeclareAffil{NRAO}{National Radio Astronomy Observatory, Charlottesville and Greenbank, USA}
\DeclareAffil{UWA}{International Centre for Radio Astronomy Research, University of Western Australia, Crawley, WA 6009, Australia}
\DeclareAffil{Berkeley}{Department of Astronomy, UC Berkeley, Berkeley, CA 94720, USA}
\DeclareAffil{RAL}{Radio Astronomy Laboratory, UC Berkeley, Berkeley, CA 94720, USA}
\DeclareAffil{BCCP}{Berkeley Center for Cosmological Physics, Berkeley, CA 94720, USA}

\affilauthorlist{
A.~P.~Beardsley\affils{ASU,UW,email},
B.~J.~Hazelton\affils{UW,UWeSci},
I.~S.~Sullivan\affils{UW},
P.~Carroll\affils{UW},
N.~Barry\affils{UW},
M.~Rahimi\affils{UMelbourne,CAASTRO},
B.~Pindor\affils{UMelbourne,CAASTRO},
C.~M.~Trott\affils{Curtin,CAASTRO},
J.~Line\affils{UMelbourne,CAASTRO},
Daniel~C.~Jacobs\affils{ASU},
M.~F.~Morales\affils{UW}, 
J.~C.~Pober\affils{Brown,UW},
G.~Bernardi\affils{SKASA,RU,CfA},
Judd~D.~Bowman\affils{ASU},
M.~P.~Busch\affils{ASU},
F.~Briggs\affils{ANU,CAASTRO},
R.~J.~Cappallo\affils{Haystack}, 
B.~E.~Corey\affils{Haystack}, 
A.~de~Oliveira-Costa\affils{MIT},
Joshua~S.~Dillon\affils{Berkeley, RAL, BCCP,MIT},
D.~Emrich\affils{Curtin},
A.~Ewall-Wice\affils{MIT},
L.~Feng\affils{MIT},
B.~M.~Gaensler\affils{USydney,CAASTRO,Dunlap}, 
R.~Goeke\affils{MIT},
L.~J.~Greenhill\affils{CfA},
J.~N.~Hewitt\affils{MIT},
N.~Hurley-Walker\affils{Curtin},
M.~Johnston-Hollitt\affils{Victoria},
D.~L.~Kaplan\affils{UWisc}, 
J.~C.~Kasper\affils{UMichigan,CfA}, 
HS Kim\affils{UMelbourne,CAASTRO},
E.~Kratzenberg\affils{Haystack}, 
E.~Lenc\affils{USydney,CAASTRO},
A.~Loeb\affils{CfA},
C.~J.~Lonsdale\affils{Haystack}, 
M.~J.~Lynch\affils{Curtin}, 
B.~McKinley\affils{UMelbourne,CAASTRO},
S.~R.~McWhirter\affils{Haystack},
D.~A.~Mitchell\affils{CASS,CAASTRO}, 
E.~Morgan\affils{MIT}, 
A.~R.~Neben\affils{MIT},
Nithyanandan~Thyagarajan\affils{ASU},
D.~Oberoi\affils{Tata}, 
A.~R.~Offringa\affils{ASTRON,CAASTRO}, 
S.~M.~Ord\affils{Curtin,CAASTRO},
S. Paul\affils{RRI},
T.~Prabu\affils{RRI}, 
P.~Procopio\affils{UMelbourne,CAASTRO},
J.~Riding\affils{UMelbourne,CAASTRO},
A.~E.~E.~Rogers\affils{Haystack}, 
A.~Roshi\affils{NRAO}, 
N.~Udaya~Shankar\affils{RRI}, 
Shiv~K.~Sethi\affils{RRI},
K.~S.~Srivani\affils{RRI}, 
R.~Subrahmanyan\affils{RRI,CAASTRO}, 
M.~Tegmark\affils{MIT},
S.~J.~Tingay\affils{Curtin,CAASTRO}, 
M.~Waterson\affils{Curtin,ANU},
R.~B.~Wayth\affils{Curtin,CAASTRO}, 
R.~L.~Webster\affils{UMelbourne,CAASTRO}, 
A.~R.~Whitney\affils{Haystack}, 
A.~Williams\affils{Curtin}, 
C.~L.~Williams\affils{MIT},
C.~Wu\affils{UWA},
J.~S.~B.~Wyithe\affils{UMelbourne,CAASTRO}
}

\begin{abstract}
The Murchison Widefield Array (MWA) has collected hundreds of hours of Epoch of 
Reionization (EoR) data and now faces the challenge of overcoming foreground and 
systematic contamination to reduce the data to a cosmological measurement. We introduce 
several novel analysis techniques such as cable reflection calibration, hyper-resolution 
gridding kernels, diffuse foreground model subtraction, and quality control methods. Each 
change to the analysis pipeline is tested against a two dimensional power spectrum figure 
of merit to demonstrate improvement. We incorporate the new techniques into a deep 
integration of 32 hours of MWA data. This data set is used to place a systematic-limited 
upper limit on the cosmological power spectrum of $\Delta^2 \leq 2.7 \times 10^4$ 
mK$^2$ at $k=0.27$ h~Mpc$^{-1}$ and $z=7.1$, consistent with other published limits, and a 
modest improvement (factor of 1.4) over previous MWA results. 
From this deep analysis we have identified a list of improvements to be made to our EoR
data analysis strategies. These improvements will be implemented in the future and detailed
in upcoming publications.
\end{abstract}
\maketitle

\section{Introduction}

Detection and characterization of the cosmic dark ages and the Epoch of Reionization 
(EoR) have the potential to inform our picture of the cosmos in ways analogous to the 
Cosmic Microwave Background (CMB) over the past several decades. The EoR in 
particular is rich with both cosmological and astrophysical dynamics as early-universe linear 
evolution gives way to non-linear structure growth and stars and galaxies reionize the 
intergalactic medium (IGM).

Several probes are being used to investigate the EoR. 
Studies of the polarization of the CMB have placed integrated constraints on 
the timing of reionization \citep[e.g.][]{Planck:2015b, Planck:2015, WMAP9}. Meanwhile, 
observations of highly redshifted quasars have placed upper bounds on redshifts by which 
reionization is complete. For example, \citet{Fan:2006a} showed that reionization must be 
complete by $z \approx 6$, a result which was confirmed independent of model by 
\citet{McGreer:2015}.
Using a $z_{\mathrm{end}}>6$ prior, the most recent analysis of \emph{Planck} data 
found a reionization redshift $z_{\mathrm{re}} = 8.8 \pm 0.9$ for a redshift-symmetric model
($z_{\mathrm{re}} = 8.5 \pm 0.9$ for redshift-asymmetric), and a duration of $\Delta z < 2.8$
\citep{Planck:2016}.
Deep optical and infrared galaxy surveys are also beginning to 
reach the redshifts necessary to further constrain reionization \citep[e.g.][]{Bouwens:2014}, 
and the James Webb Space Telescope will improve on their sensitivity 
\citep{Gardner:2006}.

The 21 cm hyperfine transition from neutral hydrogen residing in the IGM during reionization 
offers another promising method to study the EoR. Not only is the 21cm signal a 
\emph{direct} probe of the IGM, but, due to the narrow width of the transition and the 
relationship between observed frequency and line-of-sight distance, it can be used to map 
the full three dimensional space of the epoch. (For reviews, see \citealt{Furlanetto:2006}, 
\citealt{Morales:2010}, \citealt{Zaroubi:2013}, \citealt{Loeb:2013}, and \citealt{Pritchard:2012}) The first generation of instruments with primary science goals 
to detect the highly redshift 21cm signal have been built, including the GMRT (Giant Metrewave 
Radio Telescope, \citealt{Paciga:2013}), LOFAR (LOw Frequency Array\footnote{http://
www.lofar.org}, \citealt{vanHaarlem:2013, Yatawatta:2013}), PAPER (Donald C. Backer Precision Array for Probing the Epoch of 
Reionization\footnote{http://eor.berkeley.edu}, \citealt{Parsons:2010}), and the MWA 
(Murchison Widefield Array\footnote{http://www.mwatelescope.org}, \citealt{Tingay:2013, Bowman:2013}). 
Due to relatively low signal to noise, these instruments aim for a statistical measurement of 
the EoR in the form of a cosmological power spectrum. 
Complementary to power spectrum experiments are efforts to detect the sky-average global
21-cm signal from the EoR \citep[e.g.][]{Bowman:2010, Sokolowski:2015, Patra:2013, Voytek:2014}.
Meanwhile, the second generation of
21cm EoR interferometers is on its way with the Hydrogen Epoch of Reionization Array (HERA
\footnote{http://reionization.org}, \citealt{DeBoer:2016}), 
and the low frequency Square Kilometer Array (SKA1-Low \citealt{Mellema:2013}),
which will further 
refine the power spectrum measurement in early build-out stages, but will ultimately be 
capable of imaging the ionized bubbles of reionization in its later stages 
\citep{Beardsley:2015,Malloy:2013}.

Using the first generation of instruments, several upper limits have been placed on the 21 
cm EoR signal \citep{Ali:2015, Dillon:2014, Parsons:2014, Jacobs:2015, Paciga:2013}.
Furthermore, \citet{Pober:2015} and \citet{Greig:2016} were able to place constraints on physical reionization models. 
However, much work is to be done to understand the data produced by these arrays. 
Foreground subtraction and isolation has emerged as a priority in the field. Central to most 
analysis strategies is the concept of an ``EoR window" -- the region of Fourier space where 
spectrally smooth foregrounds have been isolated from the isotropic (spherically symmetric 
in Fourier space) cosmological signal \citep{Morales:2006, Bowman:2009}. More recent 
studies have shown the existence of a foreground ``wedge", the result of instrumental mode 
mixing, throwing power from spectrally smooth foregrounds to higher Fourier modes 
\citep{Thyagarajan:2015b, Thyagarajan:2015, Trott:2012, Liu:2014a, Liu:2014b, 
Hazelton:2013, Pober:2013, Thyagarajan:2013, Vedantham:2012, Morales:2012, Datta:2010}, 
while others are investigating the spectral structure of point sources themselves \cite[e.g.][]{Offringa:2016}.
The EoR window is still expected to be preserved above the wedge, and 
several analysis pipelines are under active development 
to exploit this foreground isolation
\citep[e.g. B.~J.~Hazelton et al. 2016, in 
prep; D.~A.~Mitchell et al. 2016, in prep; ][]{Jacobs:2016, Trott:2016, 
Dillon:2015b, Dillon:2013, Trott:2014}.

This paper serves two purposes: to demonstrate several new analysis techniques and their 
impact on power spectrum estimation, and to present the first deep integration 
power spectrum from the MWA. Using a three hour test set of data, we introduce several 
novel techniques including calibration of cable reflection contamination, high resolution 
gridding kernels, subtraction of a diffuse foreground model, and development of quality 
control methods which will be crucial for deeper integrations. We apply these new 
techniques to a deep integration of data from the first semester of MWA observations 
(August, 2013 -- November, 2013), with a total of 32 hours on a single EoR field and redshift range, $6.2<z<7.5$.

Our best result is a limit on the cosmological power spectrum of $\Delta^2 \leq 2.7 \times 10^4$ 
mK$^2$ at $k=0.27$ h~Mpc$^{-1}$ and $z=7.1$. This is lower than the GMRT 40 hour
limit ($\Delta^2 \le 6.15 \times 10^4$ mK$^2$ at $z=8.6$; \citealt{Paciga:2013}), but 
significantly higher than the latest PAPER result which 
integrated over 700 hours ($\Delta^2 \le 502$ mK$^2$ at $z=8.4$; \citealt{Ali:2015}).
Several EoR experiments are reaching sensitivity levels where very low systematics
become dominant. The PAPER team saw evidence that antenna cross-talk and foreground
leakage was limiting their result at several scales. The LOFAR team has seen excess noise
and diffuse foreground suppression due to calibration \citep{Patil:2016}. It is essential
for EoR experiments to understand and overcome these systematics in order to reach
a cosmological detection.

While not expecting to detect the EoR with less than a hundred hours of observations, our 
intermediate integration will serve to identify and diagnose systematics, allowing for 
improved analysis in future work. We list several directions to improve our modeling of
foregrounds and the instrumental response, which in turn will improve our calibration
and foreground subtraction. Our strategy to perform periodic deep integrations allows us
to continue uncovering systematics, refine algorithms and data analysis, and understand
the subtleties of the instrument to execute this challenging experiment.

The remainder of this paper is organized as follows. In Section~\ref{sec:MWA} we briefly 
describe the MWA instrument and the observations used in this work, in Section~
\ref{sec:analysis} we describe our analysis pipeline, in Section~\ref{sec:techniques} we 
describe several novel techniques in our analysis, in Section~\ref{sec:deep} we discuss our 
efforts to apply the analysis pipeline to a deep data integration and compare with an alternative analysis pipeline for robust estimation, and we discuss future work 
in Section~\ref{sec:discussion}, including a summary of planned improvements
to the analysis. Throughout this paper we use a $\Lambda$CDM 
cosmology with $\Omega_m=0.73$, $\Omega_\Lambda=0.27$, and $h = 0.7$, consistent 
with WMAP seven year results \citep{Komatsu:2011}. All distances and wavenumbers are in 
comoving coordinates.

\section{The Murchison Widefield Array and Observations}\label{sec:MWA}
The Murchison Widefield Array is one of several first generation radio interferometers with 
a primary science goal of detecting the 21cm EoR power spectrum. The remote Australian 
outback offers relative isolation from human-made radio frequency interference (RFI) such as 
FM radio, or TV stations. A recent study of the RFI environment at the Murchison Radio 
Observatory can be found in \citet{Offringa:2015}.

While the 21 cm EoR signal is priority, the MWA is a general observatory serving several 
science programs beyond cosmology including Galactic and extragalactic surveys, time 
domain astrophysics, solar monitoring, and ionospheric science. The array was thus 
designed with these programs in consideration and the layout was optimized using a 
pseudo-random antenna placement algorithm \citep{Beardsley:2012} to obtain a well 
behaved point spread function while retaining a dense core for EoR sensitivity 
\citep{Beardsley:2013, Bowman:2006}.\footnote{A drone flyover view of the array layout is available on YouTube at \url{https://youtu.be/yDWdTUzTUMo}.}
A full description of the science capabilities of the array is found in
\citet{Bowman:2013}.
The high imaging capability of the MWA enables us to calibrate and subtract foregrounds
based on sky models, leveraging the full capability of the observatory for the
EoR experiment. 
This strategy is in contrast to more targeted EoR experiments like PAPER, 
which utilize a highly redundant layout to enhance sensitivity. Redundant arrays can 
exploit symmetries of the instrument for quick calibration and analysis, but have poor 
point spread functions, making foregrounds more difficult to characterize.

The technical design of the MWA is reviewed in \citet{Tingay:2013}, and we highlight a 
few key characteristics that will become important in our analysis of the data. Each 
antenna of the MWA comprises 16 dual-polarization dipoles placed on a regular grid, lying 
on a ground screen. The radio frequency signals from these dipoles feed into an analog 
beamformer which uses physical delays to ``point" the antenna. The MWA contains 128 of 
these antennas, with a tightly packed core of radius 50 m, and extending out to a radius of
1.5 km for higher resolution calibration and imaging.

The beamformed signals (one for each polarization) are then transmitted to receivers in 
the field, which digitize the signal and perform a first stage coarse frequency 
channelization of the data (1.28 MHz coarse bands; \citealt{Prabu:2015}). This step 
applies a filter shape and aliases channels on the edge of the coarse bands, which will later be flagged in our analysis. 
The observer now selects 24 coarse bands (30.72 MHz total bandwidth) to pass onto the 
correlator via fiber optic link. The correlator further channelizes the data to 10 kHz 
resolution, cross multiplies signals between antennas to form visibilities, and averages in 
time and frequency to a resolution specified by the observer. For the data in this work, the 
correlator output resolutions were 0.5 seconds in time and 40 kHz in frequency. The data 
are then written to disk on a cadence of 112 seconds, constituting a single observation, or 
snapshot.

The EoR observing campaign has adopted a ``drift and shift" tracking strategy -- we point 
the telescope towards a sky field of interest, allow the field to drift overhead for about 30 
minutes until it begins to leave our field of view, then repoint the instrument at the field. The 
telescope is thus only pointed at discrete positions, or ``pointings". This tracking is 
repeated until the field is too low in the sky to track. We have identified three EoR fields 
relatively devoid of Galactic emission and bright extragalactic sources for observing: 
``EoR0" (Right Ascension (RA) $=0.00$ h, Declination (Dec) $= -27^\circ$), ``EoR1" (RA $=4.00$ h, Dec $= -27^\circ$), 
and ``EoR2" (RA $=11.33$ h, Dec $= -10^\circ$). In addition, we observe these fields in 
high and low bands centered at 182 MHz ($z\approx6.8$) and 154 MHz ($z\approx8.2$) 
respectively.

The data for this work includes two sets. First we use a set of 94 two-minute high band observations of 
the EoR0 field from August 23, 2013, which were taken early in telescope operations and 
found to be particularly well behaved. The MWA EoR collaboration has designated this set 
as a ``golden data set'', which is used to build analysis tools and compare early results 
\cite[e.g.][]{Jacobs:2016, Trott:2016, Pober:2016, Dillon:2015}. All 
techniques demonstrated here will use this golden data set. In Section~\ref{sec:deep} we 
will use a larger data set to move toward a deep integration. This data set consists of all 
EoR0 high band observations beginning August 23, 2013 (including the golden data set), 
and concluding when the field was no longer accessible for the season: when the sun was 
above the horizon for most or all of the field's transit. The final observation of this set was 
on November 29, 2013. The individual snapshots are 112 seconds of data at 0.5 second, 
40 kHz resolution. In total we analyze 2,780 snapshots, or about 86.5 hours of data.

\section{Analysis Pipeline}\label{sec:analysis}
\begin{figure}
\begin{center}
\includegraphics[width=\columnwidth]{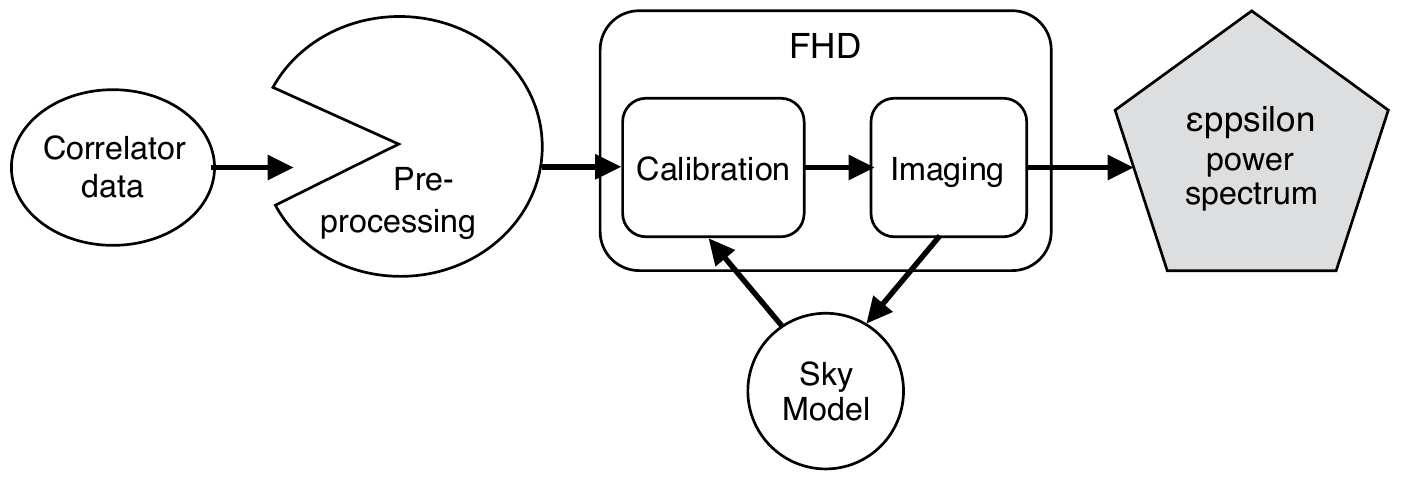}
\caption[Pipeline diagram]{
Schematic of the analysis pipeline used in this work. The raw correlator data is 
preprocessed by performing RFI and other flagging. Then the data are calibrated and 
imaged using an iterative foreground modeling approach with the FHD package. Finally, a 
power spectrum is formed using the \eppsilon package.
\label{fig:pipe}
}
\end{center}
\end{figure}

In order to ensure consistency in analysis, the international MWA EoR collaboration has 
defined two independent reference analysis pipelines. The details of this strategy are 
outlined in \citet{Jacobs:2016}. This study is based on the FHD-to-\eppsilon 
pipe, which is in turn based on the Fast Holographic Deconvolution 
(FHD\footnote{\url{https://github.com/EoRImaging/FHD}}, \citealt{Sullivan:2012}) and Error 
Propagated Power Spectrum with Interleaved Observed Noise (\eppsilon
\footnote{\url{https://github.com/EoRImaging/eppsilon}}, B.~J.~Hazelton et al. 2016, in 
prep) packages. The general flow of the pipeline is shown in Figure~\ref{fig:pipe}, and we 
give an overview below. As a point of comparison, and to demonstrate robustness of the results, we compute the output power spectra with an independent calibration and power spectrum estimation pipeline, and demonstrate consistency.

\subsection{Preprocessing}\label{subsec:preprocessing}
The preprocessing step converts the data from the non-standard memory dump format 
from the correlator's GPUs to a standard UVFITS format \citep{Greisen:2012}, 
phases the data to the center of the pointing,
performs 
flagging, averages in time and frequency, and writes to disk. For this step we use the 
\cotter package \citep{Offringa:2015}, which in turn calls the \texttt{AOFLAGGER}\footnote{\url{http://
aoflagger.sourceforge.net/}} to perform RFI flagging \citep{Offringa:2010}. 

Besides RFI, we manually flag data to remove three other effects. As mentioned in 
Section~\ref{sec:MWA}, the edges of the coarse bands contain aliasing from 
the filter applied in the digital receivers. We flag two 40 kHz channels on either side of the 
coarse band edges (total of four channels per coarse band edge). In principle the aliasing 
could be calibrated out and these channels could be recovered, but this is left for future 
work. Second, early data from the MWA also contained some instances where antennas 
were not pointed properly at the beginning of observations due to a beamformer error. This 
problem has since been resolved, but exists in our data. We therefore flag the first two 
seconds of each observation to avoid any potentially mis-pointed data. 
Finally, we flag the central 40 kHz channel of each coarse band. This channel corresponds
to the coarse band DC mode, which has been observed to contain anomalous power likely
due to very low level rounding errors in the polyphase filter bank of the digital receivers.

In the final step of preprocessing, \cotter averages the data to 2 second, 80 kHz resolution 
and writes to disk. At this stage each snapshot is contained in a single UVFITS file, which 
serves as the base unit for calibration and imaging.

\subsection{Calibration and Imaging}\label{subsec:cal_imaging}
The next step in our pipeline is to calibrate the visibilities. As shown in Figure~\ref{fig:pipe}, 
a sky model is derived from our imaged data, which feeds back into the calibration. This is 
naturally an iterative process as the sky model is refined with improved calibration 
solutions. We will describe the sky model in detail in Section~\ref{sec:techniques}, but here 
we present the production mode of analysis after the model has been determined.

Calibration is accomplished within the FHD package in two steps. First we match the raw 
data to model visibilities formed using a realistic simulation of the array response to our 
foreground model. This step results in independent complex gains for each antenna, 
80~kHz frequency channel, and polarization. In the second step we impose restrictions on these 
gains motivated by our understanding of the spectral response of the instrument. By 
reducing the number of free parameters in this step we increase our signal to noise on 
calibration solutions and avoid absorbing unmodeled confusion source flux density into our gain 
solutions which has been shown to contaminate the EoR window \citep{Barry:2016}. 
Recently, \citet{Patil:2016} demonstrated excess noise and a suppression of diffuse
foregrounds in LOFAR images after calibration and subtraction of bright discrete
foreground sources. They suggested a multi-frequency calibration solution as a potential
solution to these systematics, similar to the scheme implemented here. Indeed we have
not seen evidence of these systematics through simulations and direct propagation of
our noise \citep[B.~J.~Hazelton et al. 2016, in prep.]{Barry:2016}.

In the first calibration step we allow our complex gains to account for any direction-independent 
response on a per antenna, per 80~kHz frequency channel, per polarization basis. To 
define our gains in this way, we have made two simplifications. First we push all 
direction-dependence of the antenna response into the model of the primary beam. FHD is capable 
of using separate models for each antenna, though this has not been implemented in this 
work. The second simplification is to ignore any cross 
terms between our different axes, for example a mutual coupling between two antennas. In 
principle these terms would introduce baseline dependent gains, rather than antenna 
dependent. We have not yet seen evidence of these terms in MWA data, and so we currently 
neglect them in our solutions.

Under these assumptions we can express the measured, uncalibrated visibility for an 
antenna pair $(i,j)$ as
\begin{equation}
V'_{ij}(\nu) \approx g_i(\nu)g^*_j(\nu)V_{ij}(\nu),
\end{equation}
where $g_i(\nu)$ and $g_j(\nu)$ are the complex gains for antennas $i$ and $j$ 
respectively,
and $V_{ij}(\nu)$ is the true visibility which we aim to recover through calibration.
Because we are treating polarizations independently, we allow the antenna 
subscript to run over both East-West (E-W) and North-South (N-S) polarizations.

FHD utilizes the StEFCal algorithm described in \citet{Salvini:2014} to find the minimum 
$\chi^2$ estimate with respect to the complex gains. The result of this operation is 
estimated gains for every antenna, frequency channel, and polarization. However, with 
certain known properties of the antenna response we can reduce the number of free 
parameters in our solution. This takes us to the second step of calibration.

Our initial gain model included an amplitude bandpass common to all antennas, $B(\nu)$, 
and an antenna dependent low order polynomial in frequency. Mathematically we can 
express our restricted gain as
\begin{equation}\label{eq:cal1}
\hat{g}_i(\nu)=B(\nu)P_i(\nu),
\end{equation}
where the polynomial term can be further decomposed as
\begin{equation}
P_i(\nu) = (A_{i,0} + \nu A_{i,1} + \nu^2 A_{i,2})e^{i (\phi_{i,0} + \nu \phi_{i,1})}
\end{equation}
The coefficients $B(\nu)$, $A_{i,n}$, and $\phi_{i,n}$ are real quantities. This model allows 
us to capture arbitrary spectral response due to common antenna factors (such as the 
polyphase filter shapes or phased array response), as well as slowly varying antenna-
dependent deviations. We also found it necessary to include terms dependent on the 
length of the cable connecting the beamformers to the receivers, which we will discuss in Subsection~\ref{sec:cables}.

Once all calibration terms are found, we complete the calibration by dividing the raw data by 
the gain estimates to produce calibrated visibilities. The entire calibration process is done 
for every snapshot of EoR data, providing both a two-minute resolution time dependence of 
calibration solutions, as well as model visibilities which are used for foreground subtraction 
and diagnostic purposes.

After calibration, we form snapshot image cubes (frequency mapping to line-of-sight 
direction). The visibilities are gridded using the primary beam as the gridding kernel, placing 
the data in the ``holographic frame," which has been shown to be the optimal weighting scheme for 
combining images in the sense that it preserves all information for parameter estimation 
\citep{Morales:2009,Bhatnagar:2008}. We use a simulation-based primary beam model for MWA 
antennas developed by \citet{Sutinjo:2015} 
which incorporates an average embedded element pattern for the dipoles, as well as mutual
coupling between dipoles within an antenna.
At this stage we also average in frequency by 
a factor of two by gridding pairs of frequency channels to the same $(u,v)$ plane, resulting in 
a frequency resolution of 160 kHz. 
This is done to reduce the data volume, and has little impact on our results because
our sensitivity to the EoR is extremely low at the corresponding line-of-sight $k$ modes
\citep{Beardsley:2013}.

The data are gridded into separate cubes for even/odd interleaved time samples. 
The sum of the even and odd cubes contains power from the sky as well as noise,
while the difference between the two contains only noise power. Subtracting the power
spectrum of the difference from that of the sum yields an unbiased estimate of the sky power
\citep{Jacobs:2016}.

The gridded $(u,v)$ data are then Fourier transformed to create sky-frame images. To avoid 
aliasing we image out 90$^\circ$ from phase center (gridding resolution of a half 
wavelength), and crop the image. We found that cropping the image based on beam value 
resulted in hard edges when integrating images from different snapshots having beams pointed 
in slightly different directions. To avoid the hard edges, we predetermine a set of HEALPix 
pixels to interpolate to, and use the same set for all snapshots on a given field. The final 
cropped field of view for this work is a 21$^\circ$ square centered at RA~$=0$ h, Dec~
$=-27^\circ$. 

In principle the foregrounds could be subtracted from the data immediately after calibration, 
before gridding and imaging. However, for diagnostic reasons, we have found it beneficial to 
carry the model through the entire pipeline, and only subtract just before squaring to power 
spectrum units, allowing us to form power spectra of the dirty, model, and residual data.

FHD provides all inputs needed for the \eppsilon package, which include weights cubes (for 
proper accumulation of data), variance cubes (for error propagation), and even/odd 
interleaved data cubes (for an unbiased estimator and a direct measurement of the noise). 
We also retain both East-West and North-South polarizations. At this time we can remap 
our coordinates to cosmological units according to the relationships given in 
\citet{Morales:2004}, where angular and frequency units $(\theta_x,\theta_y,f)$ map to 
cosmological distance units $(r_x,r_y,r_z)$. Often $r_z$ is referred to as the line of sight 
dimension, and denoted as $r_{||}$. Similarly, $(r_x,r_y)$ is the plane perpendicular to the 
line of sight and is denoted as $\mathbf{r_{\perp}}$.

\subsection{Power Spectra}
In the final steps of our pipeline we integrate the snapshot image cubes, and calculate a 
power spectrum estimate. Though the integration step is formally an imaging component, 
we conceptually group it with the power spectrum step. This is because all steps of the 
pipeline up to this point have been performed on a per-snapshot basis (no communication 
between snapshots beyond foreground modeling), and the integration step can be used to 
select subsets of data for diagnosing power spectrum artifacts. The images produced by 
FHD are in the holographic frame, which is already properly weighted for combining 
images, so the integration is simply adding the cubes together, and propagating the 
weights.

The \eppsilon package performs a discrete Fourier transform (DFT) on each $r_{||}$ slice of 
the integrated HEALPix cube, forming a cube in $(\mathbf{k}_{\perp},r_{||})$, where $
\mathbf{k}_{\perp}$ represents the cosmological wavenumber in the plane perpendicular to 
our line of sight, and $r_{||}$ is the distance to the observed redshift slice along the line of 
sight. The Fourier transform along the $r_{||}$ dimension is treated separately due to 
incomplete $(u,v)$ sampling and flagged frequency channels, which leads to structure in the 
frequency sampling along any given $k_{\perp}$ pixel. We thus adopt the Lomb-Scargle 
least-squares method to determine the orthogonal eigenfunctions given our sampling 
function and estimate the total power in each $k_{||}$ mode \citep{Scargle:1982}. In 
addition, because the spectrally smooth foregrounds contain vastly more power than the 
expected EoR signal, we apply a Blackman-Harris window function prior to the $r_{||}$ to 
$k_{||}$ transform. This has the effect of trading lower effective bandwidth for higher 
dynamic range \citep[e.g.][]{Thyagarajan:2016, Thyagarajan:2013}.

The pipeline as described up to this point is applied to both the calibrated data
and the foreground model. At this point we subtract the model from the data to form
residual data as well. All three sets of data (dirty, model, and residual) are carried through 
the remaining steps to form corresponding power spectra.

After squaring and dividing by our observation window function \citep{Bowman:2006}, we 
arrive in three dimensional power spectrum space. We use the even/odd interleaved cubes 
to form a signal (odd plus even) and noise (odd minus even) power spectrum, which we 
subtract from one another to form an unbiased estimate of our signal power. This is 
mathematically equivalent to cross multiplying the even and odd cubes. The three 
dimensional power spectrum cube can next be averaged in annuli 
of constant $k_{\perp}$ and $k_{||}$ (i.e. orthogonal to the $k_{||}$ axis)
to form two dimensional 
power spectra, or spherical shells to form one dimensional power spectra where we will 
ultimately constrain the EoR.

\begin{figure}
\begin{center}
\includegraphics[width=\columnwidth]{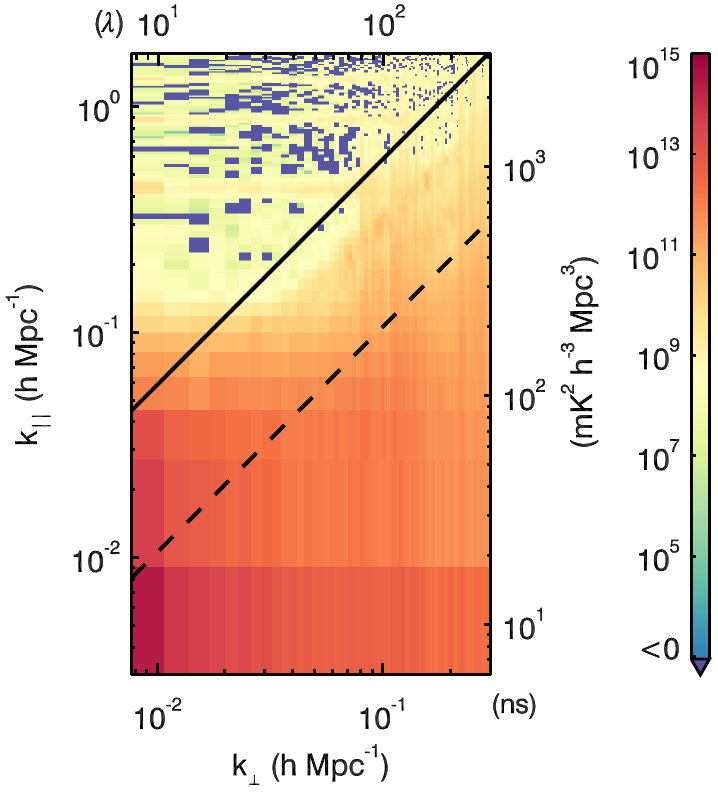}
\caption{
An example two dimensional, E-W polarization residual power spectrum formed using the 
three hour golden data set
from August 23, 2013. The foreground wedge dominates the region below the solid 
black horizon line, while regions above are mostly noise-like with the exception of horizontal 
coarse band harmonic lines. 
The vertical streaks at high $k_{||}$ and high $k_{\perp}$ are due to sparse $(u,v)$ sampling
beyond the dense core of the MWA.
The right axis shows the delay (Fourier dual to frequency) corresponding to the 
$k_||$ axis, and the top axis shows the baseline length (in wavelengths) corresponding
to the $k_{\perp}$ axis.
\label{fig:example_ps}
}
\end{center}
\end{figure}

An example two dimensional residual power spectrum formed from the three hour golden 
data set is shown in Figure~\ref{fig:example_ps}. The bulk of the residual power is in the 
so-called foreground ``wedge", indicated with the diagonal black lines where the solid line 
corresponds to sky emission at the horizon, and the dashed lines corresponds to emission 
at the edge of the MWA field of view. Above the wedge we see horizontal lines of 
contamination due to the periodic frequency sampling function. These lines are often 
referred to as the coarse band harmonics. We see vertical streaks at high $k_{||}$ and high 
$k_{\perp}$. These are due to sparse $(u,v)$ sampling beyond the dense core of the MWA 
(starting $\sim70\lambda$), 
which results in non-uniform spectral sampling after gridding. This in turn
causes foreground power to mix to high spectral modes \citep{Bowman:2009}. 
The frequency dependent sampling creates a point spread function in the $k_{||}$
for each $(u,v)$ cell. In principle covariance weighting along the frequency dimension
may be able to mitigate this leakage \citep[e.g.][]{Liu:2011}, and an \eppsilon implementation 
is under development.
We provide axes more closely related to the measurements to assist in connecting
these instrumental effects. The right axis shows delay (Fourier dual to frequency), and the
top axis shows baseline length measured in wavelengths.
Between the coarse band lines and to the left of the vertical streaks we see regions 
which contain both positive and negative (indicated by blue on the color bar) pixels. These 
regions are where we have successfully isolated the foregrounds and retained a noise-like 
EoR window in the three hour integration.

\begin{figure}
\begin{center}
\includegraphics[width=\columnwidth]{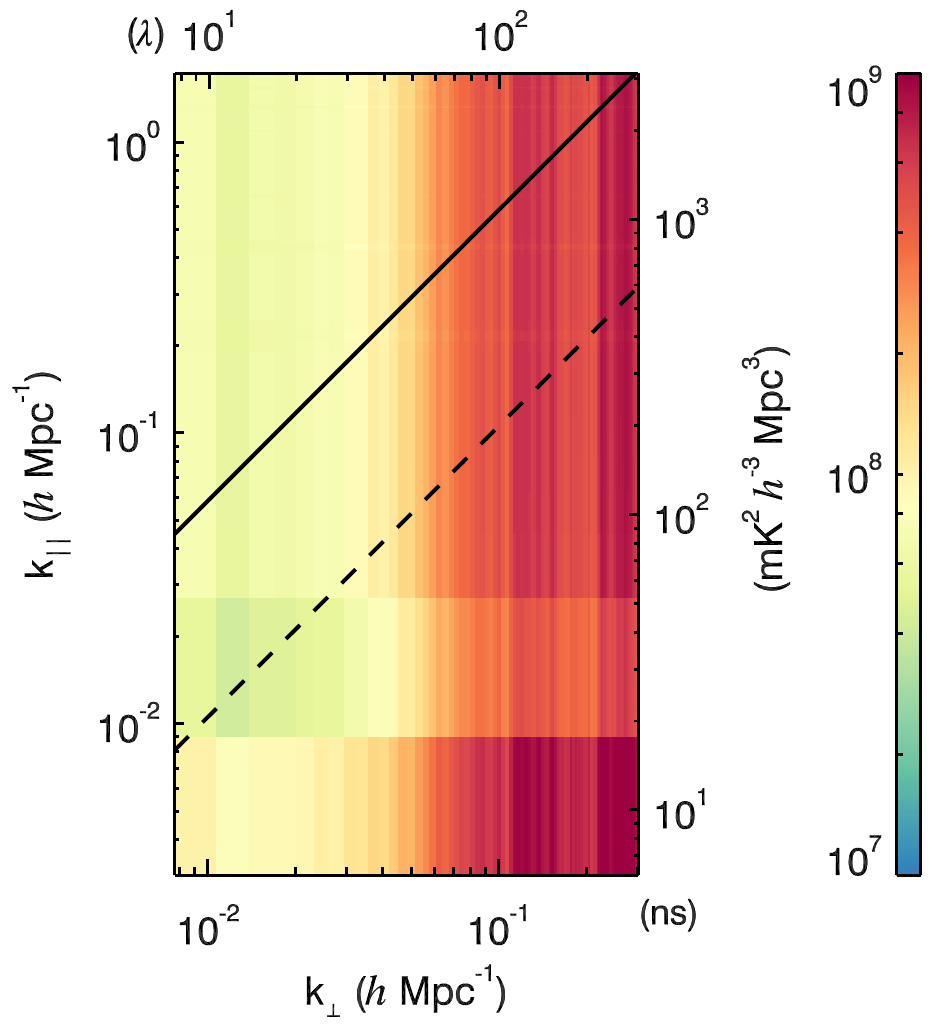}
\caption{
Uncertainty level corresponding to Figure~\ref{fig:example_ps}. The noise is relatively low at
short baseline length (low $k_{\perp}$) due to the dense MWA core. The array is more
sparse at longer baselines, and the noise increases. The second $k_{||}$ bin has slightly
lower noise than the rest of the $k_{||}$ modes due to the interaction between the
Lomb-Scargle periodogram with the Blackman-Harris window function applied to the data.
\label{fig:example_2derror}
}
\end{center}
\end{figure}

Figure~\ref{fig:example_2derror} shows the uncertainty level corresponding to the two 
dimensional power spectrum in Figure~\ref{fig:example_ps}. These levels
are found by propagating noise measured from calibrated visibilities through the entire
pipeline. The dense core of the MWA results in relatively low noise at low $k_\perp$.
As the $(u,v)$ density of the array decreases at longer baselines, the noise increases.
The noise is mostly constant across $k_{||}$ because the thermal noise amplitude is
relatively constant over frequency and the Fourier transform distributes the noise power
across all modes \citep{Morales:2004}.
The exception is the second $k_{||}$ bin, which has slightly lower noise than the rest of the 
$k_{||}$ modes. 
As explained in B.~J.~Hazelton et al. (in prep.) this is a result of the Lomb-Scargle periodogram 
interacting with the Blackman-Harris window function we apply to the data. Most of the bins 
have the noise equally distributed between cosine and sine terms, with the exception of the 
$k_{||}=0$ bin where the cosine term carries all the information. Due to the correlations 
introduced by the Blackman-Harris window function the first non-zero $k_{||}$ mode is highly 
covariant with the bottom bin -- therefore the noise on this mode's cosine term is lower than 
other higher order cosines.

\section{Analysis Methods for Testing Data Quality}\label{sec:techniques}

The two-dimensional power spectrum is a useful figure of merit (FoM) as we improve and 
refine our analysis pipeline. Foregrounds and systematics often manifest with characteristic 
shapes in this space, enabling us to diagnose problems and quantify improvements 
\citep{Morales:2012}. We use the MWA 3 hour ``golden'' data set from August 23, 2013, to 
repeatedly form power spectra to test and refine our analysis. While an exhaustive catalog 
of these improvements is outside the scope of this paper, we 
demonstrate the utility of the 2D power spectrum as a FoM with a few key techniques we 
have employed in our analysis. 

\subsection{Cable Dependent Calibration}\label{sec:cables}

Early in our analysis it became apparent that the bandpass of each MWA antenna requires 
more free parameters than those described in Equation~\ref{eq:cal1}. In particular, differing 
lengths of cable connecting the beamformers to the receivers
 lead to different bandpass shapes due to signal attenuation. 
We therefore allow the amplitude bandpass factor to depend on cable length, $B_{\alpha}
(\nu)$, where $\alpha$ denotes the cable type.\footnote{For logistical purposes the 
beamformer to receiver cables were installed in six set lengths. Cables of length 90, 150, 
and 230 meters are RG-6, while 320, 400, and 524 meter cables are LMR400-75.}

\begin{figure}
\begin{center}
\subfigure{\includegraphics[width=\columnwidth]{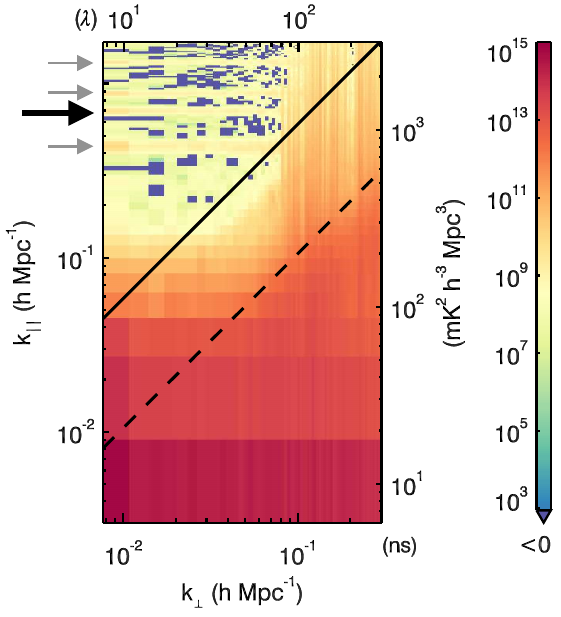}}
~
\subfigure{\includegraphics[width=\columnwidth]{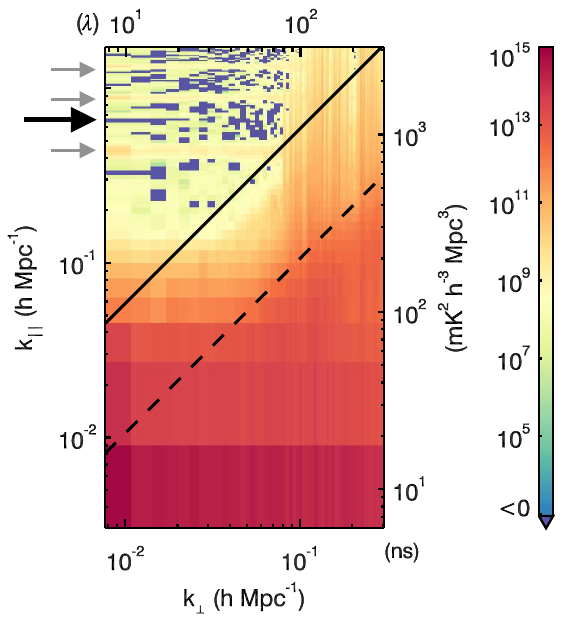}}
\caption{
\emph{Top:} Single polarization dirty power spectrum formed from the golden data set, 
before implementing cable reflection fitting into the calibration loop. 
The small gray arrows point to the bands resulting from the periodic coarse band sampling.
The black arrow points 
to a horizontal band at $k_{||} \approx 0.7$~h~Mpc$^{-1}$, which cannot be accounted for 
by the coarse bands. \emph{Bottom:} After we implement the cable reflection fitting in our 
calibration solutions, we see the reflection line disappears.
\label{fig:cables}
}
\end{center}
\end{figure}

The top panel of Figure~\ref{fig:cables} shows a 2D power spectrum using the gain model 
described so far. Three of the horizontal lines in the EoR window can be attributed to the 
coarse band gaps, as they reside at the harmonics expected for 1.28~MHz periodic 
sampling. But the sampling cannot account for the fourth line, highlighted by the arrow. The 
corresponding delay time of the line, $\tau \approx 1.23$~$\mu$s, corresponds almost 
exactly to twice the signal travel time through the 150 meter cables (with velocity 0.81c). We 
therefore introduced a reflection term into our gain model, which allows our restricted gains 
to account for the interference of the incident signal with a round-trip reflected signal in the 
cable. 
A similar approach was taken at lower frequencies in \citet{Ewall-Wice:2016}.
The full gain model is expressed as
\begin{equation}\label{eq:cal2}
\hat{g}_i(\nu)=B_{\alpha}(\nu)\left(P_i(\nu)+R_i(\nu)\right),
\end{equation}
where the reflection term can be further decomposed as
\begin{equation}
R_i(\nu) = R_{i,0} e^{-2\pi i \tau_{i} \nu}.
\end{equation}
The reflection coefficient, $R_{i,0}$ is allowed to be complex, while the reflection delay, 
$\tau_i$ is real. After fitting for all other parameters, we fit for the reflection mode. 
We found that if we did not first fit and remove the polynomial terms, our reflection
fits would be dominated by the large power in the smooth structure, resulting in solutions that
did not match physical reality.
The reflection fitting is
done for antennas with suspected reflections, the most offensive of which is seen in the 
antennas with cables of 150 meters. Because the cables were not cut at exact lengths 
(variations on order tens of centimeters), we solve for both the reflection coefficient and 
delay by performing a direct Fourier transform to a highly over resolved delay grid and 
selecting the mode where the reflection amplitude is largest. 
This fitting produces a single reflection coefficient for each antenna with nominal cable
length of 150 meters.

The resulting 2D power spectrum after fitting for cable reflections is shown in the bottom 
panel of Figure~\ref{fig:cables}, where the reflection line is suppressed below the noise 
level. This example demonstrates the power of the 2D power spectrum as a figure of merit. 
The power in the reflection line was about five orders of magnitude below the foregrounds, 
making it very difficult to detect in image-based metrics. However the power spectrum is 
specifically designed to be sensitive to low-level spectral structure, and using the two 
dimensional spectrum allows us to identify the ``shape'' of contamination, enabling a 
precise diagnosis.

\subsection{Gridding Kernel Resolution}
\begin{figure}
\begin{center}
\subfigure{\includegraphics[width=\columnwidth]{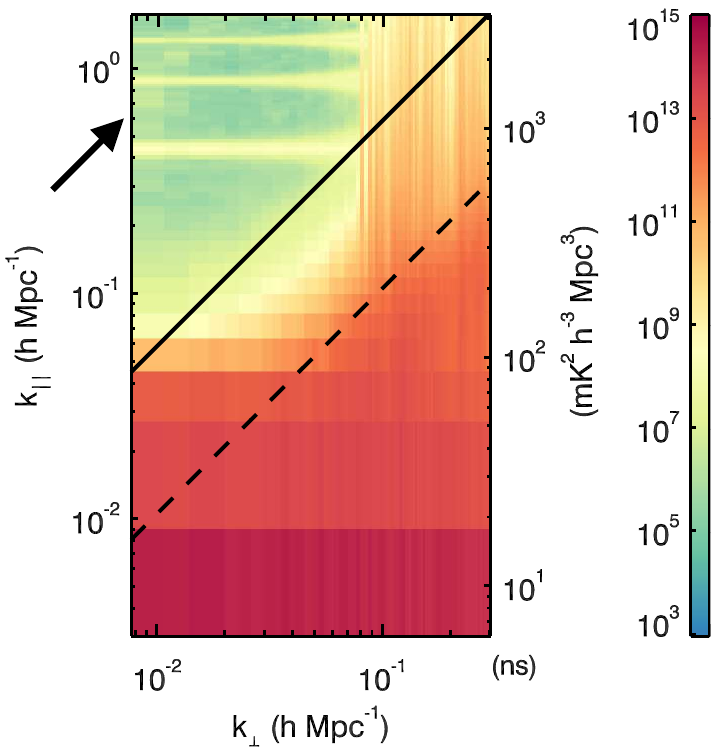}}
~
\subfigure{\includegraphics[width=\columnwidth]{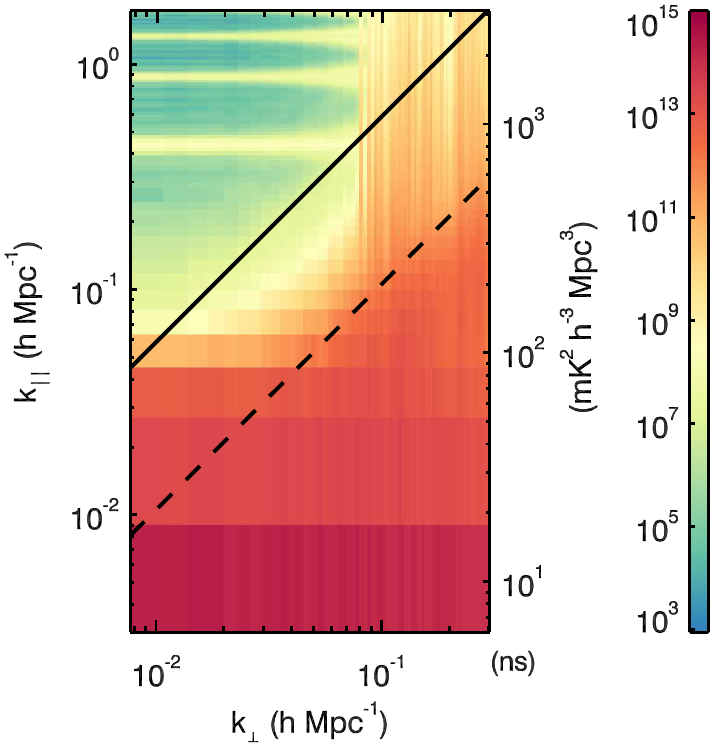}}
\caption{
\emph{Top:} Single polarization model power spectrum demonstrating the contamination in 
the EoR window when using insufficient gridding kernel resolution. The window has a floor 
comparable to an expected EoR signal, and a faint super-horizon line appears at high 
$k_{||}$. 
The black arrow indicates the location and direction of the faint line.
\emph{Bottom:} Model power spectrum after increasing the gridding kernel 
resolution from 0.04 wavelengths to 0.007 wavelengths. The floor is now far below the 
expected EoR signal level. 
\label{fig:beam_res}
}
\end{center}
\end{figure}

A similar, low-level effect which had the potential to contaminate the EoR window is shown 
in Figure~\ref{fig:beam_res}. The spectra shown are the \emph{model} spectra -- calculated 
by propagating the sky model visibilities through the entire pipeline. Despite our foreground 
model not containing spectral structure, the EoR window in the top panel seems to have a 
floor at a level $\sim 10^7$~mK$^2$~h$^{-3}$~Mpc$^3$, comparable to predicted EoR 
signals. In addition there is a faint line in the upper-left of the plot which has the same slope 
as the wedge, but seems to originate far beyond the horizon.

The source of this floor and super-horizon line was traced to the resolution at which we 
formed the gridding kernel when gridding visibilities. In the interest of computational speed 
and efficiency, the kernel is pre-calculated at a high resolution, then a nearest neighbor 
lookup table is used to approximate the beam values at discrete pixels. 
This is a common practice in most analysis software packages.
The kernel 
resolution used in the top panel of Figure~\ref{fig:beam_res} was 0.04 wavelengths -- much 
smaller than the half wavelength grid corresponding to horizon to horizon imaging. 
However, as a baseline migrates in $(u,v)$ coordinates across frequencies, it 
undergoes discrete steps between kernel values used. This effectively results in small 
baseline position errors which shift periodically in frequency, resulting in power being mixed 
into the window and a relatively strong harmonic at the frequency of the shifting (the faint 
line in the upper left of the window). While we illustrate this effect with a model power 
spectrum, the same problem exists in the data, but is below the noise levels at three hours 
of integration.

This effect is distinct from more fundamental wedge effects 
\citep[e.g.][]{Morales:2012, Hazelton:2013} in that it is a result of the computational limitation
of the analysis. The primary contribution to the foreground mode mixing is due to
the loss of information as each baseline samples a range of $(u,v)$ modes. On the other
hand, the effect discussed here is due to small positional offsets in the gridding kernel
and only exists after gridding.

We resolved this issue by increasing the resolution at which we form the kernel. While the 
most accurate answer is to model at infinite resolution, it is not computationally feasible to do 
so. Instead we chose a resolution at which the effect no longer materially 
impacts the power spectrum. With experimentation we found a kernel resolution of 0.007 
wavelengths was sufficient. The resulting model power spectrum is shown in the bottom 
panel of Figure~\ref{fig:beam_res}. 
The contamination within the EoR window now drops significantly lower. Of course, without
knowing the exact level of the cosmological signal, we cannot know if this level is sufficiently 
low. This effect may need to be revisited if the EoR power spectrum is lower than 
predicted.
The improvement in beam
model resolution came at the cost of 20\% increase in memory usage for our imaging
pipeline.

\subsection{Improving Point Source Model}

Our pipeline uses two modes of FHD -- the full deconvolution mode to identify point sources 
and build a catalog, and a production mode where we calibrate using the sky model and 
subtract it from the data without further fitting. Because full deconvolution on every 
observation from the MWA is computationally not feasible, we currently restrict ourselves to 
the golden data set to build our model. The process of building the catalog is presented in 
\citet{Carroll:2016}. They applied machine learning classification methods 
to select reliable detections from the full deconvolution FHD mode, culminating in the KGS 
catalog\footnote{KGS is an abbreviation of KATALOGSS, the KDD (Knowledge 
Discovery in Databases) Astrometry, Trueness, and Apparent Luminosity of Galaxies in 
Snapshot Surveys}. The work here uses an early iteration of the KGS catalog that was available at the 
time of analysis.

It has been shown that subtracting a foreground model strictly within the main lobe of the 
primary beam will not be sufficient to suppress the power spectrum wedge and unlock the 
EoR window \citep{Thyagarajan:2015b, Thyagarajan:2015, Pober:2016}. 
We are in the process of repeating the model building described above using additional
MWA observations pointed away from our field of interest.
Until that work is complete we supplement the extent of 
our point source model using additional catalogs. We accomplish this through a hierarchical 
catalog pulling from our early KGS catalog, the MWA Commissioning Survey\footnote{The 
more complete GLEAM Survey \citep[][N.~Hurley-Walker et al., in review]
{Wayth:2015} was not available at the time of analysis.} \citep[MWACS, ][]
{Hurley-Walker:2014}, the Culgoora catalog \citep{Slee:1977}, and the Molonglo Reference Catalog 
\citep[MRC, ][]{Large:1981}. Source flux densities are prioritized in this order based on our 
confidence in their predicted flux density at 182 MHz. We first cluster the source lists to avoid 
redundant sources, using a 3.5 arcmin neighborhood radius, then select the flux density from the 
highest priority catalog. Spectral indices were obtained from the MWACS and Culgoora 
catalogs when available, otherwise a two-point spectral index was estimated for Culgoora-
MRC matches or MRC-SUMSS matches. All other sources were given a spectral index of 
$-0.8$, the previously determined median spectral index of sources below 1.4 GHz
\citep{Oort:1988, Hunstead:1991, DeBreuck:2000, Mauch:2003}.
The spectral index was used to extrapolate the flux density from the catalog frequency to 
182 MHz, but a uniform spectral index of $-0.8$ is used within our band when forming 
model visibilities for calibration and foreground subtraction.

\begin{figure*}
\begin{center}
\includegraphics[width=\textwidth]{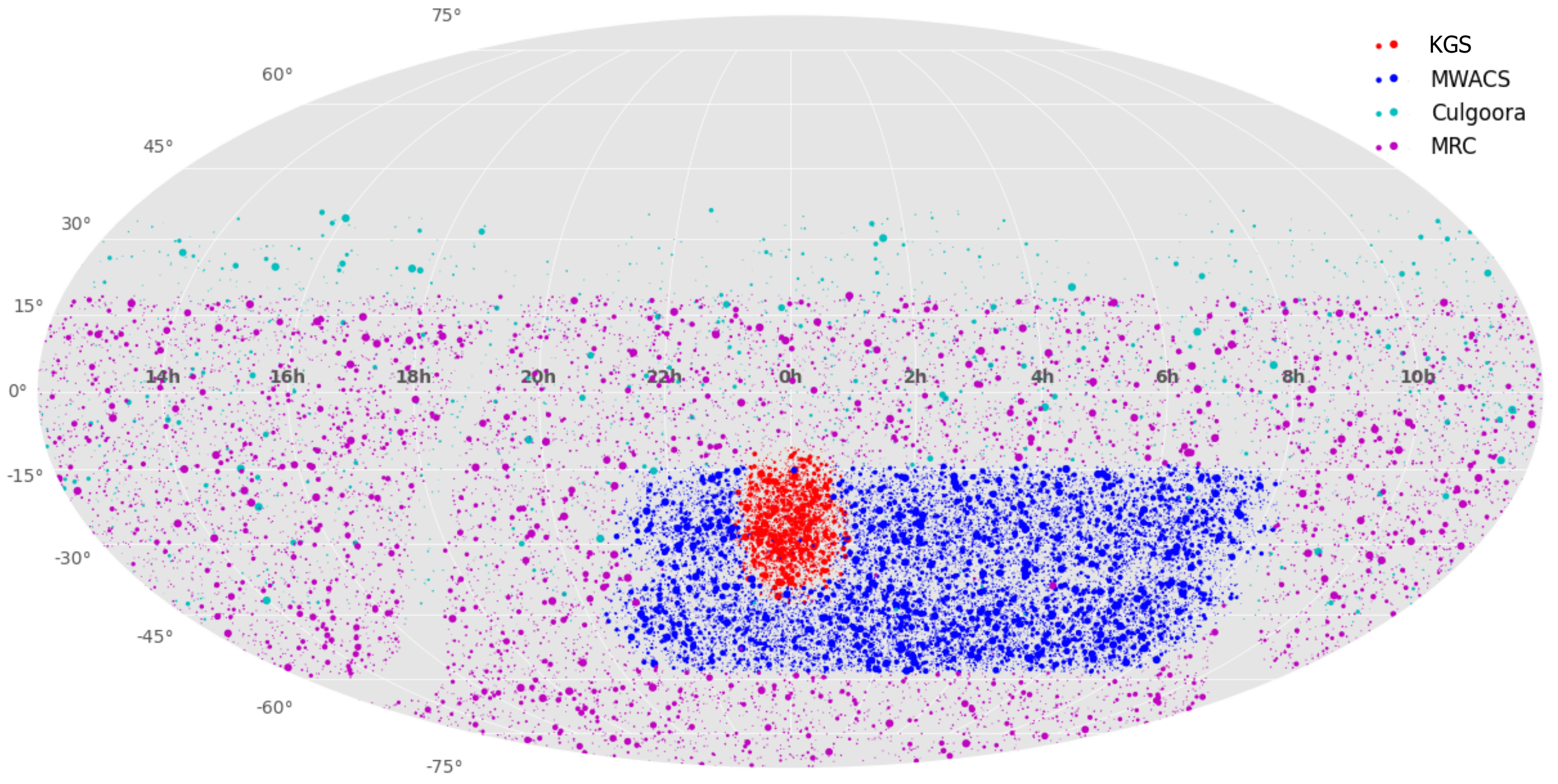}
\caption{
Hierarchical catalog used for foreground subtraction. This catalog combines the source lists 
from the KGS catalog, the MWACS catalog, Culgoora sources, and the MRC, 
prioritizing in that order. The EoR0 field corresponds to the red KGS patch, while we use the 
other catalogs to fill in the sidelobes of the MWA. The size of each dot is proportional to the 
182~MHz flux density of the source, clipped at 20~Jy.
\label{fig:master_catalog}
}
\end{center}
\end{figure*}

The resulting hierarchical catalog is shown in Figure~\ref{fig:master_catalog}. The EoR0 
field corresponds to the red KGS source patch. All sources above the horizon and within a 
primary beam value greater than 1\% maximum (including sidelobes) are included in both 
our calibration and foreground subtraction models.  

To show the effect our hierarchical catalog has on the power spectrum figure of merit, we 
ran our entire analysis pipeline on the golden data set with two foreground models -- first 
with only the MWACS catalog, and again with the full hierarchical catalog. We then 
compare the resulting 2D power spectra to inspect whether the new catalog results in more 
accurate calibration and foreground subtraction.

\begin{figure}
\includegraphics[width=\columnwidth]{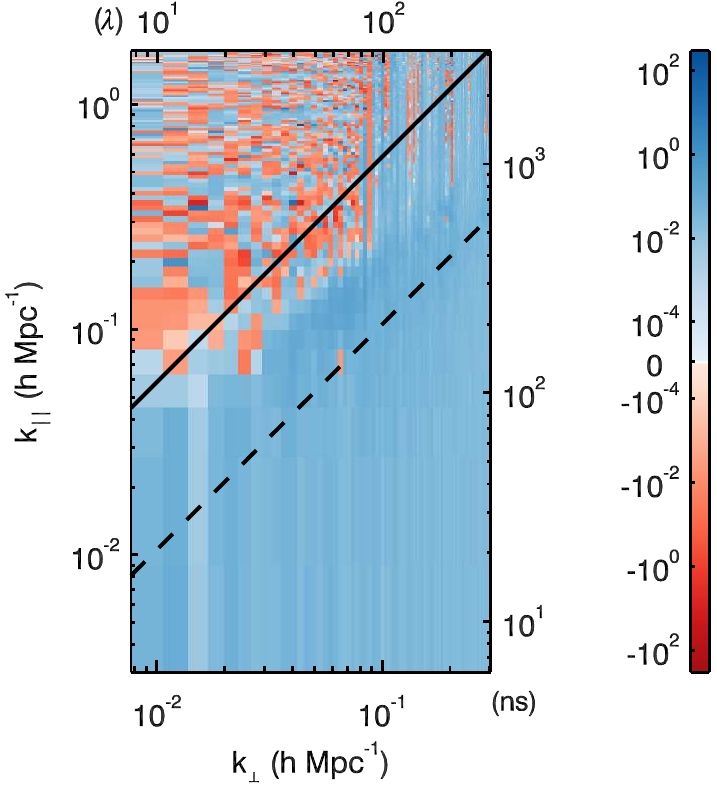}
\caption{Power spectrum ratio difference, according to Equation~\ref{eq:diff_ratio}. This 
metric shows the difference in the fractional E-W residual 2D power spectrum, comparing 
the MWACS catalog and our hierarchical catalog. By dividing the residual spectrum by the 
corresponding dirty spectrum before subtracting we remove the effect of the overall flux 
scale change due to differing calibration solutions. We see that the wedge is completely 
positive (higher fractional residual when using the MWACS catalog), confirming that the 
hierarchical catalog subtracts more power from the data. This also reassures that the 
calibration using the new catalog is more accurate. 
\label{fig:mwacs_ratio}
}
\end{figure}

Because we use the input foreground model to calibrate our visibilities, the gain estimates 
for our two runs differed. In particular we saw the overall flux scale of the calibrated 
visibilities was higher when using the full hierarchical catalog,
owing to a more complete sky model requiring lower amplitude gains to describe the data. 
This difference is on order a part in $10^3$ in mK$^2$ units, but is largely amplified
when observing the difference in the power spectra.
In order to put both residual 
2D power spectra onto the same scale for comparison, we first divide each 
$(k_{\perp},k_{||})$ cell
by the 
corresponding dirty power spectrum pixel, then subtract to arrive at a ratio difference. 
\begin{equation}\label{eq:diff_ratio}
\mathtt{ratio\;difference = \frac{residual_1}{dirty_1} - \frac{residual_2}{dirty_2}}
\end{equation}
Here we use subscript 1 to represent the MWACS catalog spectra, and 2 for the 
hierarchical catalog spectra. The ratio difference is shown in Figure~\ref{fig:mwacs_ratio}. 
The entire wedge being positive (blue) demonstrates that using the new catalog resulted in 
successfully subtracting higher fractional power. Not only does this indicate we subtract 
more power, but it also confirms that our calibrated data is more closely matched to the 
model, meaning that our calibration solutions in general are more accurate.

J.~L.~B.~Line, et al (2016, in review) performed similar analysis, but with an emphasis on the
positional accuracy of catalog entries. By using simulations of visibilities they were able to 
isolate the effect of using perfect versus offset source positions. In their simulations source 
position offsets of $14\%$ of the synthesized
beam-width had significant impact on foreground subtraction, confirming the
importance of a complete and accurate point source catalog in order to retain a clean EoR
window.

\subsection{Diffuse Foreground Model Subtraction}

In addition to a point source foreground model, we also introduce a diffuse emission model 
within the primary field of view of the MWA. Though the EoR0 field was chosen to be 
relatively devoid of Galactic emission, we find this diffuse structure still highly contaminating 
due to the very high sensitivity of the MWA at large scales. For computational purposes we 
have divided the diffuse emission into two regimes -- the faint clouds in our main lobe, and 
the bright plane and other structures in the sidelobes. While the latter has been studied by 
many people and a Global Sky Model (GSM) is readily available 
\citep{deOliveira-Costa:2008}, the computational obstacle of simulating the instrument 
response to a full sky diffuse model is yet prohibitive, though under active pursuit 
\citep{Thyagarajan:2015}. We instead focus here on diffuse structure within the primary field 
and leave the full-sky model to future iterations of analysis.

This first iteration of modeling the diffuse structure within our primary field was done by 
simply using output point-source subtracted residual images of FHD from the three hour 
golden data set. We combine the images from three hours to leverage the rotation of the 
earth to improve the point spread function (PSF) of the instrument. This integrating of 
images is done with uniform weighted data to minimize the effect of double instrument 
convolution (once when the data are taken, again when we use the output as a model). We 
then form a pseudo Stokes I image by adding beam-weighted East-West and North-South 
polarizations. We also integrate in frequency to form a single continuum image for our 
model. By forcing our model to contain no spectral structure we mitigate the risk of 
subtracting cosmological signal. 

Future iterations of this model will contain a spectral index 
and multiple polarization components.
\citet{Lenc:2016} demonstrated the presence of strong polarized large-scale structures 
in the EoR0 field at 154 MHz with varying Faraday depths. They observed rotations in the
Stokes Q--U plane due to ionospheric conditions, which will need to be taken into account
in our future diffuse models.

Ultimately we need model visibilities to subtract from our data. However, rather than storing 
a model for all observations (different pointings and phase centers), we find it most simple 
to treat the model as a single HEALPix image which can be used to create model visibilities 
for each independent observation. FHD treats each pixel in the image as a point source at 
the pixel location with total flux density equal to the surface brightness of the diffuse image times the area 
of the pixel and creates model visibilities for each snapshot in the same way that it imports 
a catalog of point sources. This is similar to the strategy employed by 
\citet{Thyagarajan:2015} to model diffuse structure.

\begin{figure}
\begin{center}
\includegraphics[width=\columnwidth]{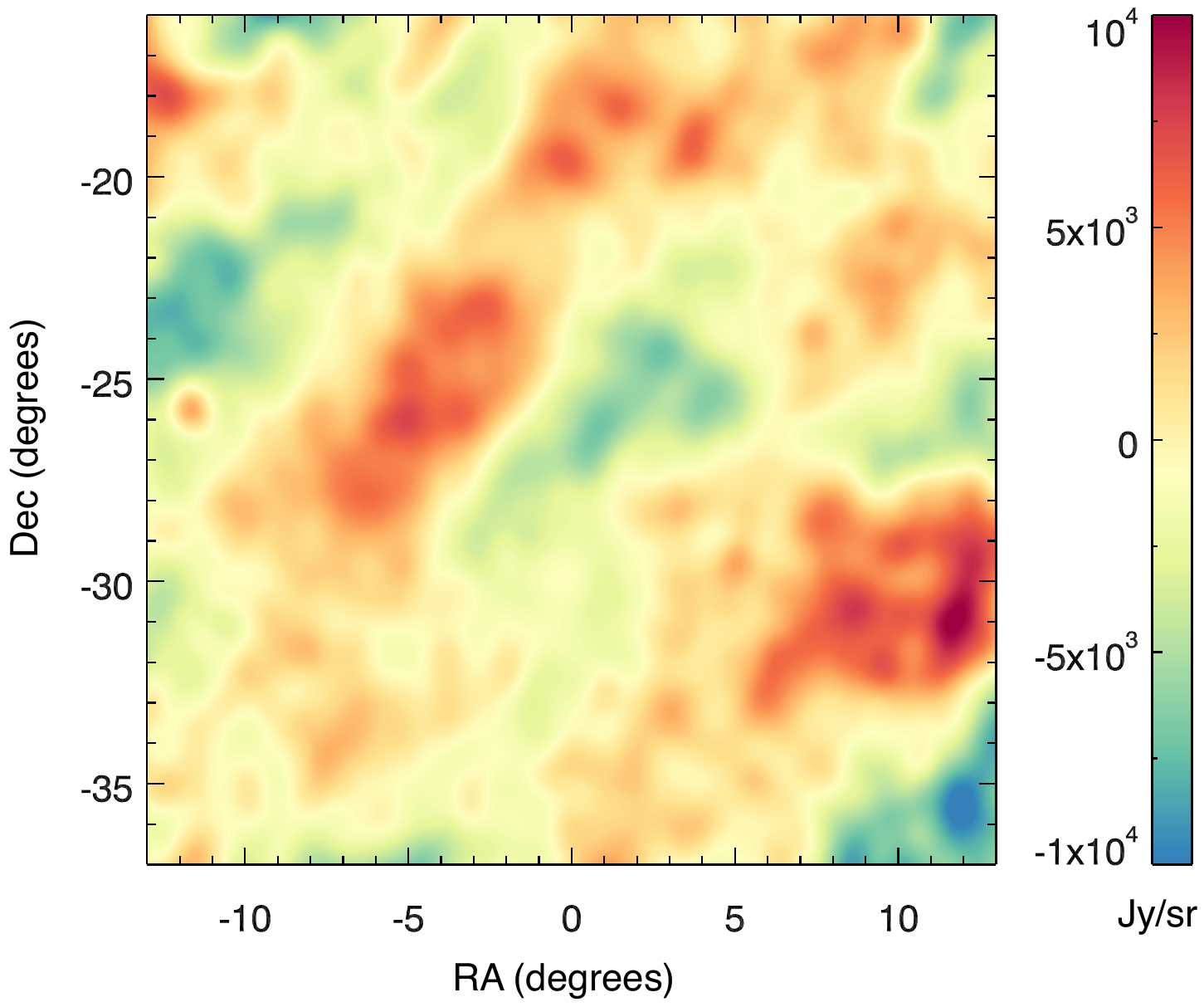}
\caption[Diffuse model used for subtraction]{
The diffuse foreground model within the EoR0 field used for foreground subtraction. This 
model was created using residual images from the golden data set. The image shown is 
smoothed to degree scales to emphasize the large scale structure and approximate an 
MWA snapshot natural weighting. Note that the negative brightness values are a 
consequence of not sampling the zero-spacing part of the $(u,v)$ plane.
\label{fig:diffuse}
}
\end{center}
\end{figure}

The diffuse model used for this work is shown in Figure~\ref{fig:diffuse}. While the actual 
model image is uniform weighted with resolution $\sim\!6$~arcminutes, we show it 
smoothed to degree scales to emphasize the large scale structure and approximate what 
the MWA instrument observes with a natural weighting.

For this iteration of analysis we only use the diffuse model for subtraction, and omit it for 
calibration. At the time of writing short baselines were not producing reliable results in the 
calibration loop. Instead we chose to omit the diffuse model and mask baselines shorter 
than 50 wavelengths for calibration purposes. However, we add the diffuse model and 
unmask short baselines when performing foreground subtraction.
We compare the total residual power in the images by squaring and summing
the residual image cube from before and after the diffuse subtraction. 
This includes all scales measured by the instrument, but because the image cubes are
naturally weighted, the $\mathbf{k}$ space weighting is the inverse of 
Figure~\ref{fig:example_2derror}, meaning $k_\perp \lesssim 0.05$ h Mpc$^{-1}$ dominates the
average. We saw a 70\% reduction in residual power when including our diffuse model
 -- a strong indication that our diffuse model improves our 
subtraction. Future iterations will incorporate the model in calibration and gradually improve 
the model itself. 

\begin{figure}
\begin{center}
\subfigure{\includegraphics[width=\columnwidth]{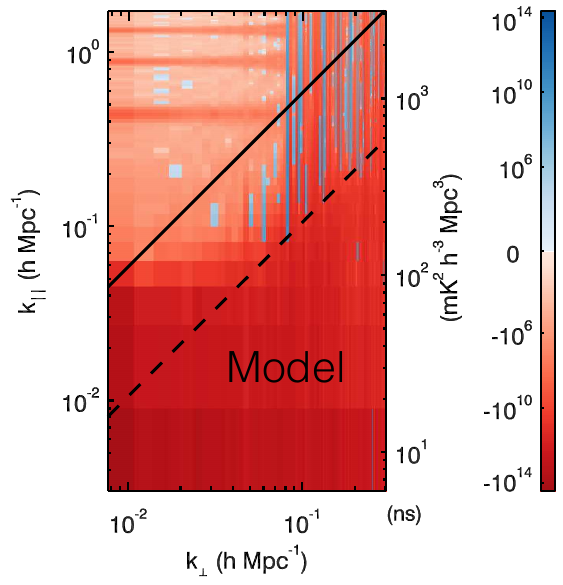}}
~
\subfigure{\includegraphics[width=\columnwidth]{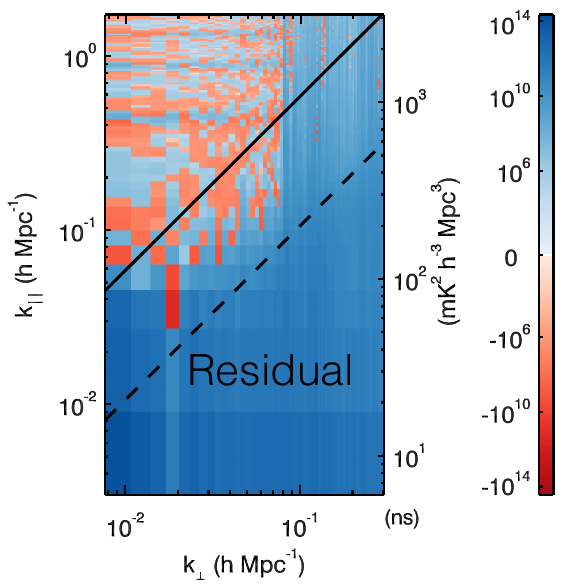}}
\caption{
Single polarization (E-W) power spectrum differences from the golden data set showing the 
effect of subtracting a diffuse foreground model. The calibration is identical with and without 
the diffuse model, so the dirty power spectra are identical. 
In the top panel the model power spectrum difference is almost entirely negative 
(red), indicating that the diffuse model added a large amount of power to the foreground 
model. The bottom panel shows the residual power spectrum difference is positive in the 
wedge (blue), demonstrating a successful subtraction.
While this first attempt at a diffuse model was rudimentary, it 
successfully removed 70\% of the residual power. 
\label{fig:diffuse_diff}
}
\end{center}
\end{figure}

We demonstrate the impact of our diffuse model again by running the golden data set 
through our entire analysis pipeline with and without using the diffuse model. Because the 
foreground models used for calibration in the two runs are identical (diffuse is omitted for 
calibration), the calibration solutions were identical and therefore the dirty power spectra 
were identical as well. We compare the model and residual power spectra by direct 
subtraction in Figure~\ref{fig:diffuse_diff}. 
The top panel shows our point source foreground model power spectrum
minus the model power spectrum when including diffuse. 
Because our diffuse model added power to the full foreground model, the entire plot is
negative (red).
Even the EoR window is red (albeit at a much lower level than the wedge) because of power
leakage from non-uniform spectral sampling and from the dynamic range limit of the 
Blackman-Harris window function.
The bottom panel is the difference in residual 
power. The wedge is completely positive (blue), indicating that the diffuse model 
successfully subtracted from the dirty visibilities. The differences in the window are positive 
and negative, indicating that 
the differences in the foreground model in this region are below the three hour noise level.

\section{A Deep Integration}\label{sec:deep}
Up to this point we have only discussed testing and analysis on the three hour golden data 
set. We next turn towards a deeper integration, incorporating the techniques described above. 
We start with all MWA observations taken on the EoR0 field at 182~MHz between August 
23, 2013 and November 29, 2013. This includes 2,780 snapshots, or about 86.5 hours of 
data. We first make data quality cuts, then form power spectra.

\subsection{Data Selection}\label{sec:selection}
All 2,780 snapshots in the data set are preprocessed, calibrated, and imaged using the 
pipeline described in the previous sections of this article. Similar to tests on the golden data 
set, we rely heavily on the 2D power spectrum as a diagnostic tool, this time to identify and 
excise poor quality data.

\begin{figure*}
\begin{center}
\includegraphics[width=\textwidth]{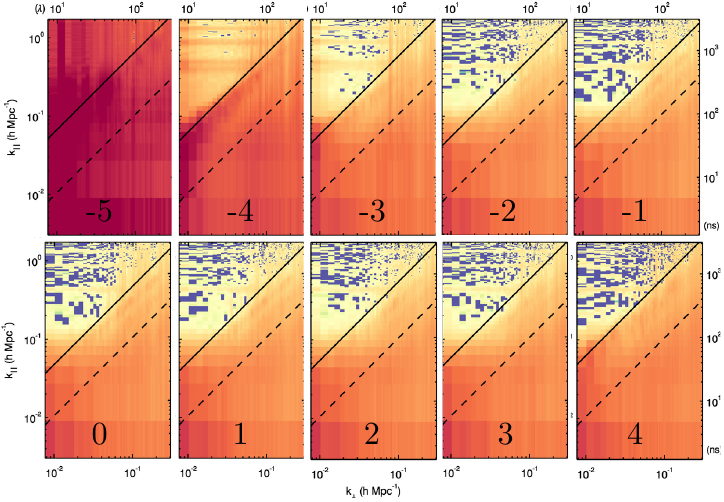}
\caption[Per pointing jackknife]{An example jackknife test. For this test we divided the data 
into days and pointings. This is an example array of power spectra (residual E-W 
polarization) for a single day, August 26, 2013. The early pointings are heavily contaminated 
by the Galaxy in the sidelobes, the window becomes more clear near zenith, and we can 
see trace contamination at the end of the night (pointing +4) when the Galaxy has risen 
again.
\label{fig:jackknife}
}
\end{center}
\end{figure*}

\begin{figure}
\begin{center}
\includegraphics[width=\columnwidth]{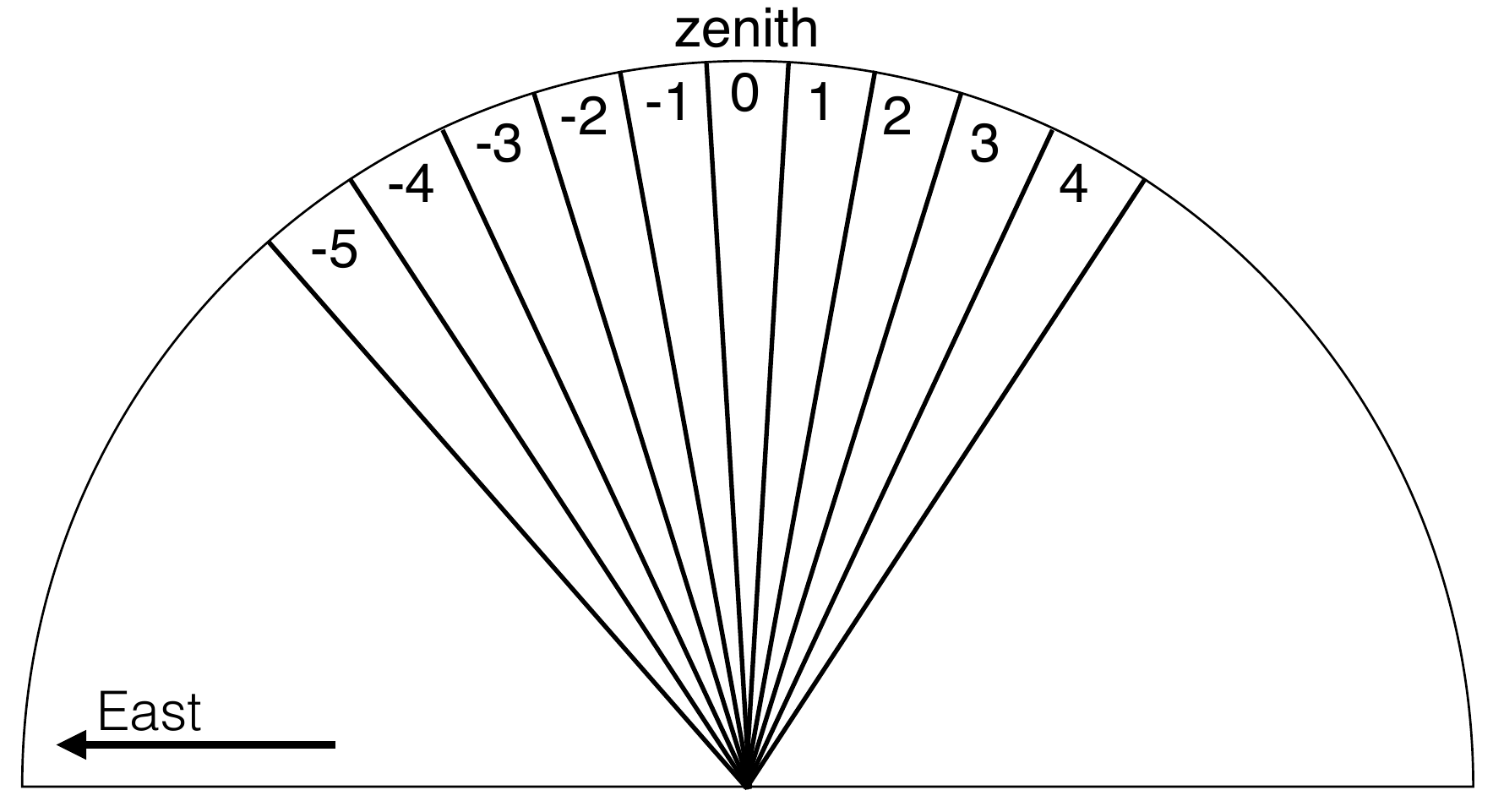}
\caption{This is a cartoon depiction of the numbering scheme used to label pointings in
Figures~\ref{fig:jackknife} and \ref{fig:wedge_cut}. Looking South, a field transits from the
East (left), and the telescope begins observing about 5 pointings before zenith (-5). As the
field drifts overhead, we periodically repoint the telescope to recenter the field. Each shift
increments the pointing label by one, and the zenith pointing is defined as zero.
While a +5 pointing (and beyond) is possible with the MWA, this observing campaign
did not contain any such data because we instead switched to the EoR1 field at that time.
\label{fig:pointing}
}
\end{center}
\end{figure}

We use the jackknife method to filter out bad data and detect patterns. This involves dividing
the data into many subgroups and forming power spectra. The goal is to find observational
parameters which affect the quality of the data, such as the day or time of night.
A powerful grouping is to divide the data into the observation day 
and ``pointing" (which direction the antennas were pointed while tracking the EoR0 field). 
We show an example of one day's worth of per-pointing power spectra in Figure~
\ref{fig:jackknife}. The pointings are labeled sequentially with $-5$ corresponding to five 
pointing steps prior to zenith transit, $0$ corresponding to zenith, and $+4$ corresponding 
to four steps after zenith. 
This numbering scheme is shown graphically in Figure~\ref{fig:pointing}.
Early in the night ($-5$ through $-3$) the bulk of the Galactic plane 
was still above the horizon, and despite being very far from our primary field of view, highly 
contaminated our observations through the sidelobes of the instrument. We then have 
relatively well-behaved pointings, with the EoR window dominated by noise, until the final 
pointing of the night when we can see evidence of the Galaxy rising again indicated by 
strong lines of power at the edge of the foreground wedge. We saw the same contamination 
on all days of observation and decided to cut all $-5$, $-4$, $-3$, and $+4$ pointings from 
our data set. In principle it may be possible to model the Galactic plane well enough to 
account for its presence, but we leave this to future work.

Motivated by the manual classification done with power spectrum jackknives, we developed 
another metric to quickly predict power spectrum quality for each snapshot in our data set. 
We do this by forming a delay spectrum \citep{Parsons:2012b} from the raw, uncalibrated 
visibilities. We then calculate an estimated total EoR window power by integrating the 
power above the horizon line and below the first coarse band line. While we would ideally 
use calibrated visibilities for this metric, we have found that the uncalibrated power is 
strongly correlated to the calibrated power, and the cuts we make below are independent of 
calibration. This is encouraging because future analysis could use this metric before 
processing the snapshots, saving valuable computing resources\footnote{Because
calibration involves generating model visibilities for each snapshot, approximately half
the computational cost of the entire imaging pipeline is required to obtain 
calibration solutions.}.

\begin{figure}
\begin{center}
\subfigure[]{\includegraphics[width=\columnwidth]{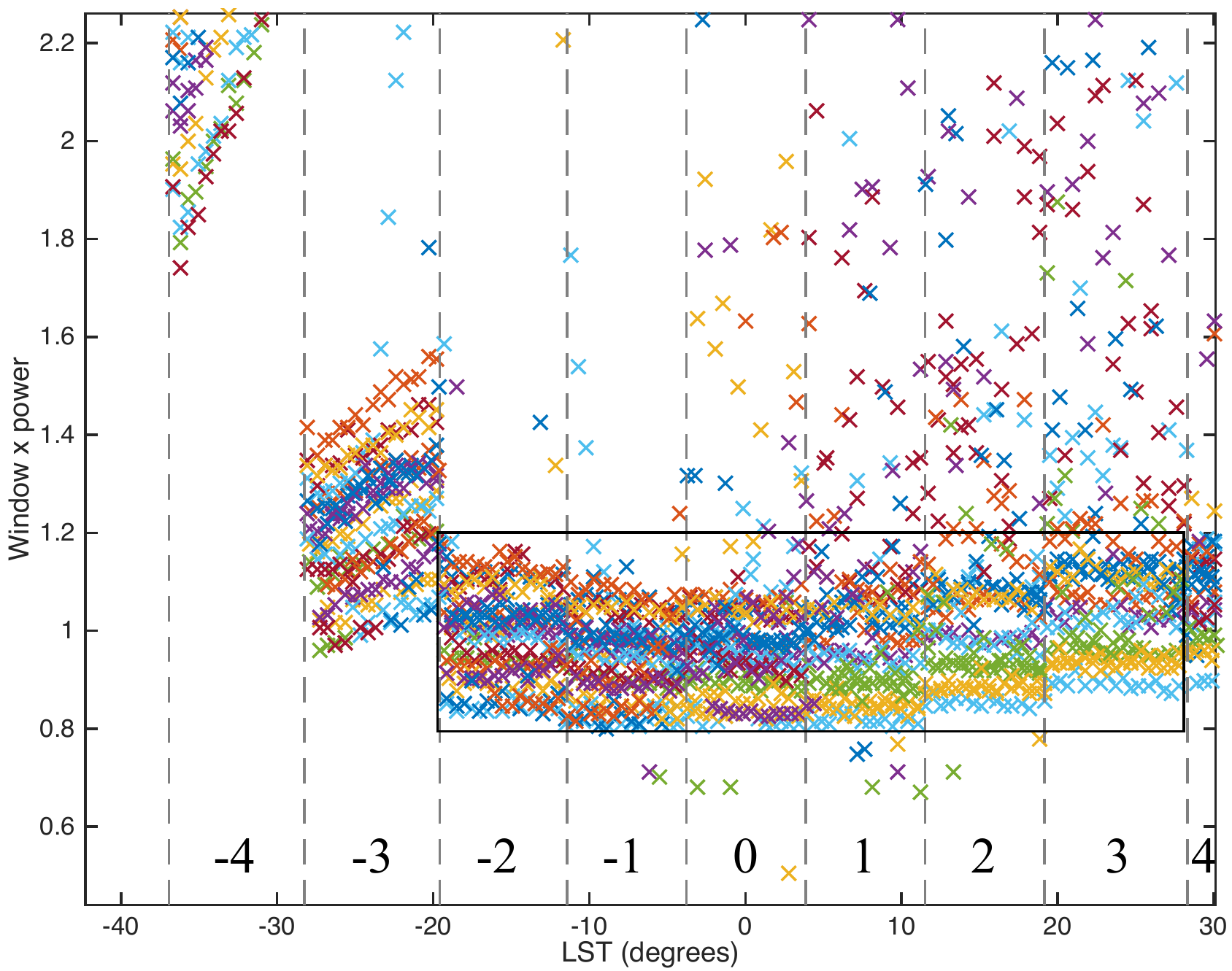}\label{fig:wedge_cut1}}
~
\subfigure[]{\includegraphics[width=\columnwidth]{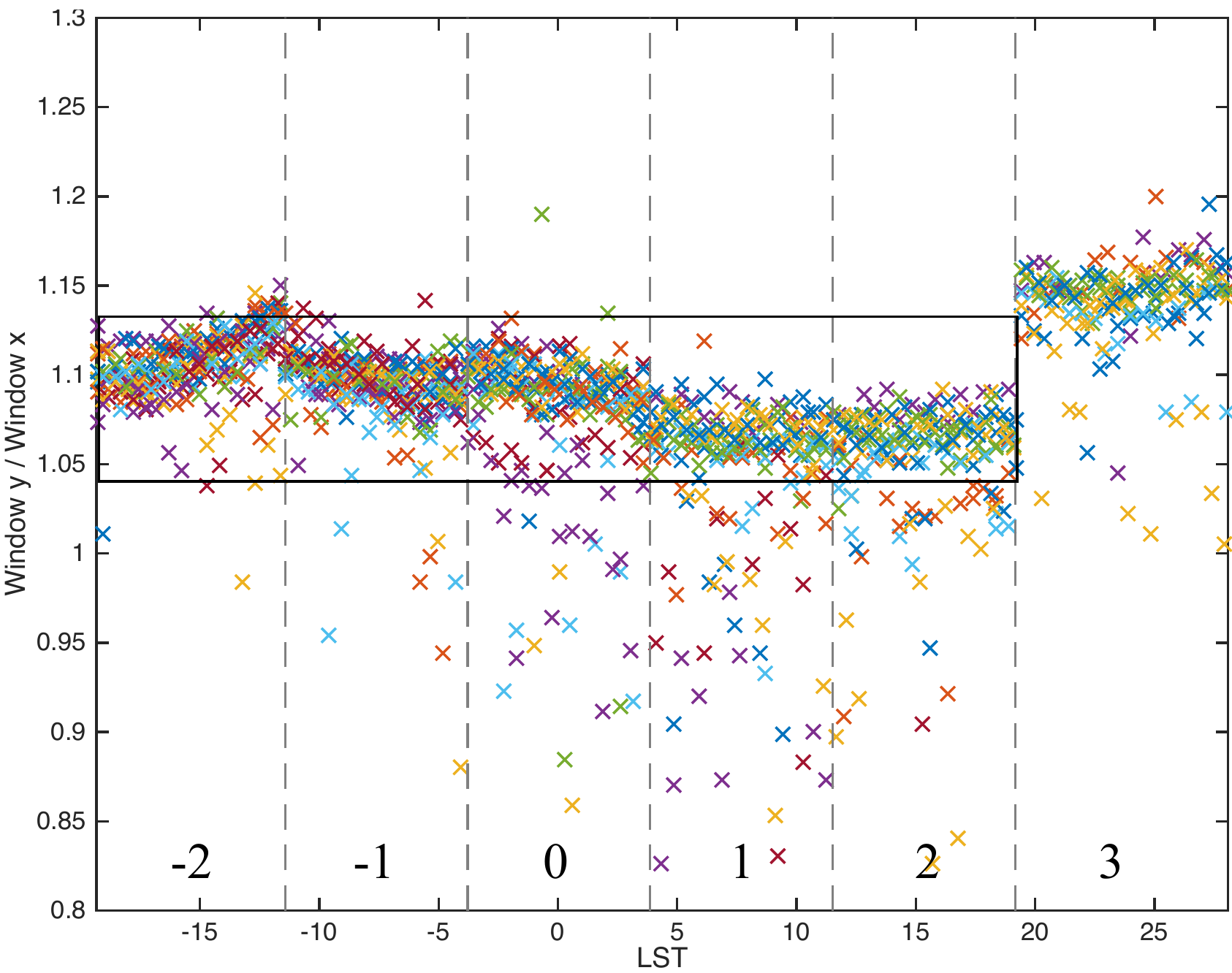}\label{fig:wedge_cut2}}
\caption[Window power cut]{
\emph{Top:} Window power for each observation in our data set. The window power is 
calculated from the delay spectrum of uncalibrated data as a fast quality metric. 
Because uncalibrated data is used, the units are arbitrary.
Each color 
is a distinct day of observing, and the vertical dashed lines represent pointing shifts, with 
the pointing numbers indicated at the bottom of the plot above the horizontal axis. The black 
box indicates observations which passed this cut. \emph{Bottom:} Data cut based on 
window power polarization ratio. For each snapshot that passed the total power cut, we plot 
the ratio of window powers in the N-S and E-W polarizations. Clear outliers can be seen, 
including the whole of pointing +3. The black box again indicates the selected data.
\label{fig:wedge_cut}
}
\end{center}
\end{figure}

We plot the window power for each snapshot in our data set in Figure~\ref{fig:wedge_cut1}. 
We first note that our conclusion from the jackknife tests is confirmed here -- early pointings 
contain strong contamination and have high window power. We also see many outlier 
snapshots with excess power. Inspecting power spectra from these individual snapshots 
confirmed poor data quality, even after calibration. The cause of these contaminants is yet 
unknown, but could easily be attributed to low-level RFI that was missed by the 
\texttt{AOFLAGGER}, or intermittent hardware failures. To remove the poor data we made 
the cut shown with the black box, keeping only snapshots inside.

Next we compare the power in our two instrumental polarizations. When conducting our 
jackknife tests we saw contamination could exist in one polarization and not the other -- 
potentially due to RFI, or an instrumental failure. We plot the ratio of N-S power to E-W 
power in Figure~\ref{fig:wedge_cut2}, after applying the previous cut. We see the ratio is 
generally flat, with the exception of some outliers and almost all of pointing $+3$. We 
suspect that the excess N-S power in the later pointing is due to the Galaxy leaking back 
into the sidelobe of the N-S polarization, but not in the other. Again, we make the cut shown 
with a black box.

Our final data cut was made when inspecting the snapshot residual continuum images output from 
FHD. 
We estimate a proxy for the residual flux density from foreground sources by calculating
the RMS of the fractional residual flux densities for all subtracted sources greater than
0.5 Jy within half-beam power. The fractional residual flux density is measured as the ratio
of the residual image pixel value (Jy) at the source position to the integrated flux density
of the source.
We found
that most snapshots had residual flux density RMS $<10\%$ with some outliers,
which we cut from the data. This cut largely overlapped with previous cuts, but removed an 
additional 95 observations. The cause of the high residual flux density RMS is yet unknown and is 
not localized by observation day, LST, pointing, or any other jackknife we have performed.

After the cuts described above, we are left with 1,029 snapshots, or just over 32 hours of 
data. This is an aggressive cut, with the aim of removing all poor quality data and retaining 
a data set as clean as we can determine from our data quality checks. As mentioned earlier, more sophisticated analysis could allow us to 
recover data cut from this study in the future (e.g. modeling and removing the Galaxy in 
observations pointing far off zenith).

\subsection{Results}\label{sec:results}

In this section we discuss the results after processing the remaining 32~hours of data 
through the \eppsilon pipeline, and place an upper limit on the EoR power spectrum.

The bandwidth of the MWA and our data set is 30.72 MHz, but in order to avoid the effects 
of cosmic evolution over the span of our measured redshift range, we limit observations to 
about 8 MHz, or $\Delta z \sim 0.3$ at our frequency. We do this by dividing the band into 
three overlapping sub-bands of 15.36 MHz each, which, after applying the Blackman-Harris 
window function, will result in effective bandwidths of 7.68 MHz. We label these sub-bands 
as ``low" ($z\approx 7.1$), ``mid" ($z\approx 6.8$), and ``high" ($z\approx 6.5$).

\begin{figure*}
\begin{center}
\includegraphics[width=\textwidth]{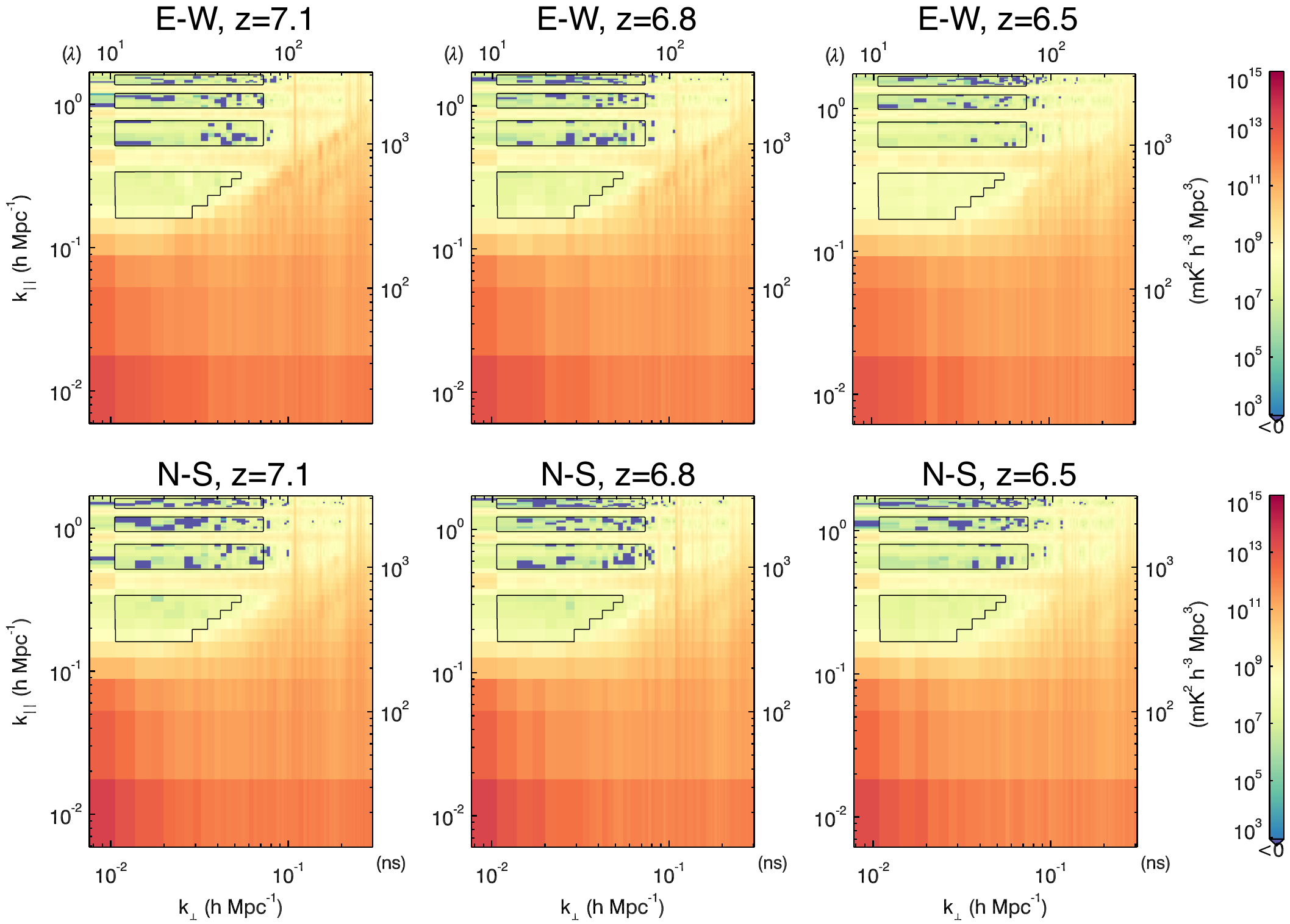}
\caption[Deep 2D sub-band power spectra]{
Residual two dimensional power spectra for the three sub-bands used to place limits on the 
cosmological signal. The left panels show the low band, centered at 174.7 MHz, or a 
redshift of 7.1. The middle panels show the mid band, centered at 182.4 MHz, or a redshift 
of 6.8. The right panels show the high band, centered at 190.1 MHz, or a redshift of 6.5. 
The top row corresponds to East-West instrumental polarization, while the bottom row 
shows North-South polarization.
The contours show the window used for averaging to one dimension.
\label{fig:2d_ps_subband}
}
\end{center}
\end{figure*}

The 32 hour integrated residual power spectra for our three sub-bands and both 
instrumental polarizations are shown in Figure~\ref{fig:2d_ps_subband}. A number of 
features can be seen in these spectra. In all cases the foreground wedge is prominent, with 
extra power at large scales (low $k_{\perp}$). This is an indication that our diffuse model needs to 
improve for deeper subtraction. 

The lowest region of the EoR window, above the horizon line and around $k_{||} \approx 
0.25$ h Mpc$^{-1}$, contains purely positive bins, indicating non-noise-like power. The 
cause for this seeming leakage is yet unknown, and limits our integration in this 
analysis. One suspected origin of this contamination was individual snapshots with high 
contamination that slipped through our cuts. However, dividing the data into random 
subsets and comparing results indicated that there is not a small number of offending 
observations, instead the leakage appears to exist at a low level in all the data.

Another potential cause of the foreground leakage into the EoR window is insufficient
calibration quality -- particularly the spectral shape of the instrumental bandpass.
Investigations are underway to combine snapshot observations for higher signal-to-noise
calibration solutions, as well as more sophisticated parameterization of the gain model in
Equation~\ref{eq:cal2} to account for more physical effects such as dependence on
ambient temperature variation.

Moving up in the window, between the first two coarse band harmonic lines we can see an 
additional faint line. This is the re-emergence of the 150 meter cable reflections. While the 
method described in Section~\ref{subsec:cal_imaging} sufficiently calibrated out this 
reflection line for the three hour golden set integration, the lower noise level in this deeper 
integration shows it is not completely removed. Future analysis will require higher 
signal-to-noise on the reflection fitting, which may require combining snapshots for calibration 
solutions. 

Between the coarse band harmonic lines and the reflection line, we do see regions where 
our spectra appear noise-like (positive and negative values). This is encouraging despite 
the leakage at lower $k_{||}$, and motivates us to be selective when binning to one 
dimensional power spectra.

\begin{figure*}
\begin{center}
\includegraphics[width=\textwidth]{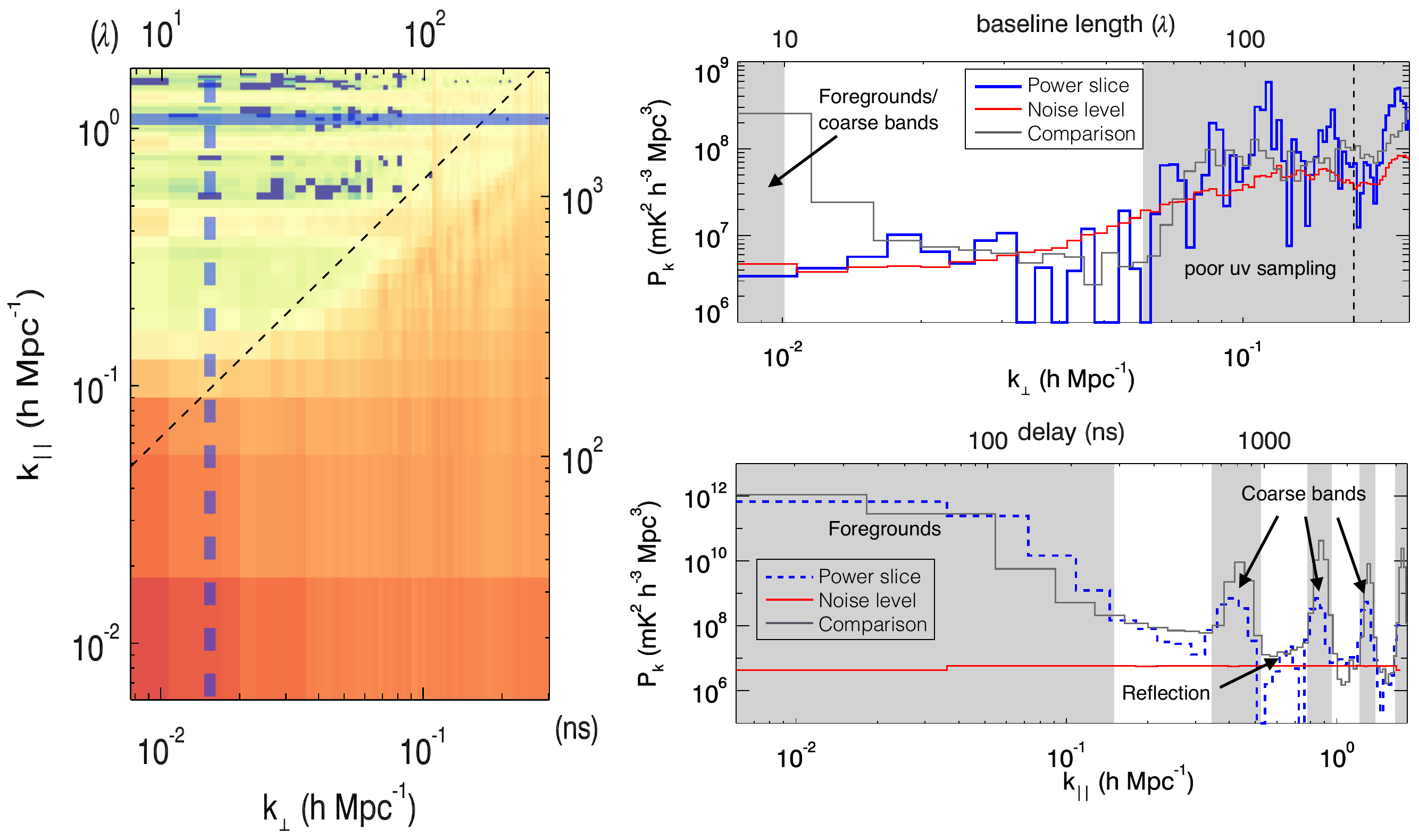}
\caption{
\emph{Left:} The N-S, $z=6.8$ two dimensional power spectrum repeated from 
Figure~\ref{fig:2d_ps_subband}. Here we have superimposed blue lines to show slices in
$k_{\perp}$ (solid) and $k_{||}$ (dashed). \emph{Top right:} Residual power as a function
of $k_{\perp}$ for fixed $k_{||}$, averaged in rings from the full 3D power cube (blue). The gray boxes show regions which will be excluded from
the one dimensional averages. The red line shows the 1-$\sigma$ noise level, and
the vertical black dashed line shows the intersection of this slice with the wedge. 
\emph{Bottom right:} Residual power as a function of $k_{||}$ for fixed $k_{\perp}$, again averaged in rings from full 3D power cube (blue). The strong
foreground contamination is evident at low $k_{||}$, and the coarse band harmonic lines
appear as expected. The thin gray line in both right panels is the equivalent power slice from the RTS$+$CHIPS comparison pipeline, which is discussed in Section~\ref{subsec:ref_pipe}.
\label{fig:slices}
}
\end{center}
\end{figure*}

Figure~\ref{fig:slices} shows slices of power that help isolate the contamination. The two 
dimensional power spectrum is the N-S, mid band spectrum from 
Figure~\ref{fig:2d_ps_subband}. We have drawn a horizontal solid blue line to indicate the 
slice used to plot power as a function of $k_{\perp}$ for a fixed $k_{||}$ (top right), and a 
vertical dashed line to show the slice used to plot power as a function of $k_{||}$ for a fixed 
$k_{\perp}$ (bottom right). 
In the $k_{\perp}$ power plot we can see leakage at high $k_{\perp}$ from the residual of the 
vertical streaks due to poor sampling at large $(u,v)$. While the increased noise would
mean that we will down weight the bins at large $k_{\perp}$, the contamination is even
larger, meaning it would bias our estimate even though those bins are down weighted.
The low end of $k_{\perp}$ contains exceptionally bright foregrounds at large scales, which,
due to the width of the coarse band harmonics, contaminate most of the $k_{||}$ modes. 
With both these regimes in consideration, 
we exclude bins with $u<10\lambda$ and $u>70\lambda$, shown with gray boxes.
Between these two regimes are many bins consistent with the noise level, which we include
in our one dimensional averages.

Similarly we inspect the $k_{||}$ power in the bottom right plot of Figure~\ref{fig:slices}. Here
we see the huge contamination from foregrounds at low $k_{||}$. While the leakage drops 
significantly, our measurements do not reach the noise level until after the first coarse
band harmonic. However, we chose to include bins below the first coarse band with 
$k_{||}>0.15$ h Mpc$^{-1}$ because the cosmological signal is expected to contain 
substantially higher power at large scales, and ultimately these low $k_{||}$ bins provide
our most competitive limits despite being systematic limited. We exclude the coarse band
harmonics by masking out the harmonic bin itself and two bins on either side (total of five
bins per harmonic).

The final mask we apply is the wedge. We found that a small buffer beyond the horizon was
necessary to completely mask out the wedge, consistent with \citet{Dillon:2015}. We
implement the buffer by increasing the slope of the horizon line by 14\%.
This line is shown as the
dashed diagonal line in the left panel of Figure~\ref{fig:slices}.

The various masking described above is summarized with the black contours shown on the
two dimensional power spectra in Figure~\ref{fig:2d_ps_subband}. The contours show
the cuts in two dimensional space, but the masking is actually performed directly in the
three dimensional power spectrum cube and averaged directly to one dimension.

\begin{figure*}
\begin{center}
\includegraphics[width=\textwidth]{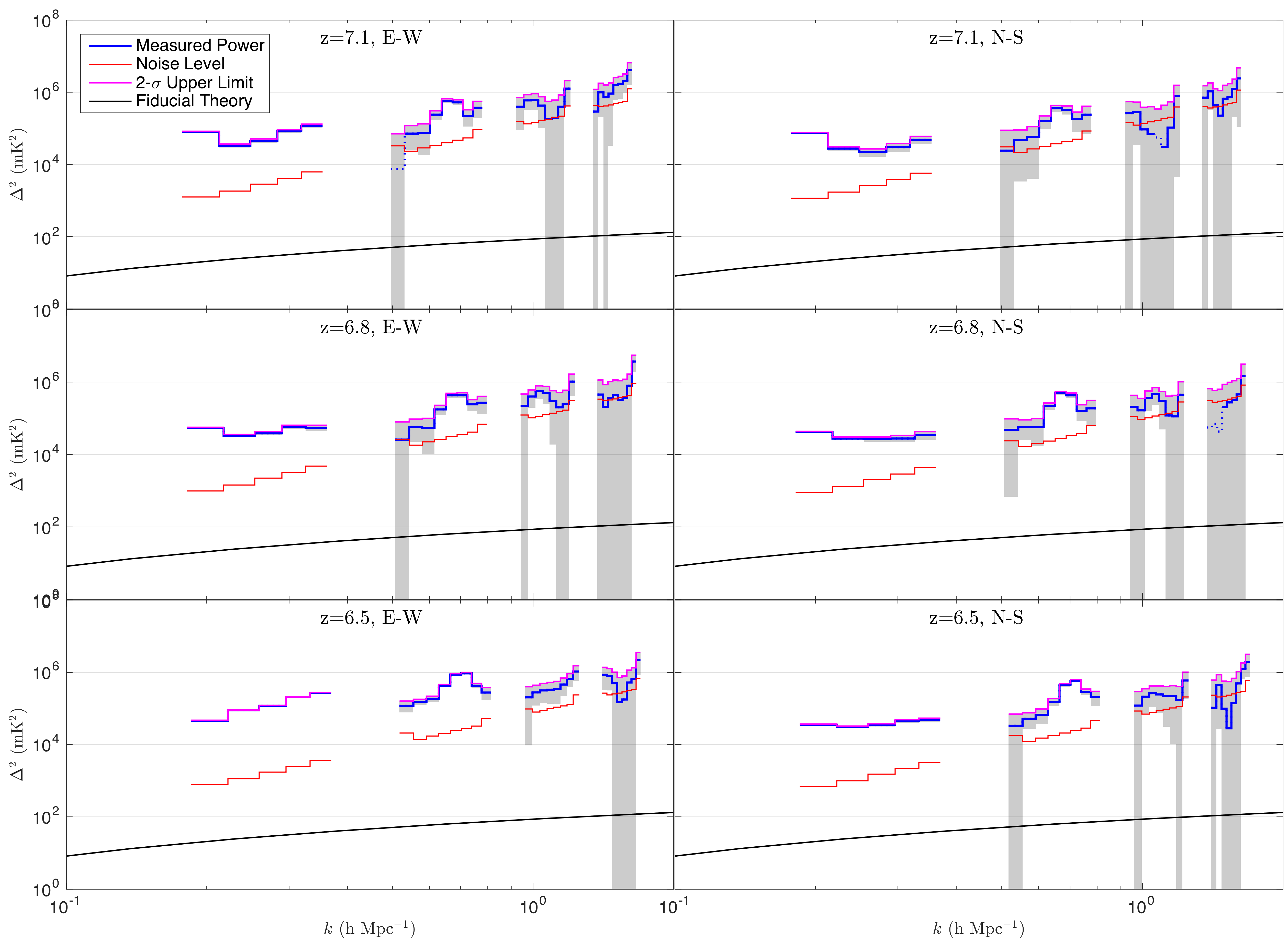}
\caption[1D deep power spectra]{
One dimensional power spectra for our three sub-bands and both instrumental 
polarizations. The solid blue line shows the measured power spectrum with step widths 
corresponding to the bin size used in the average. Where the measured signal is negative 
we plot the absolute value with a dotted line. 
The gray boxes show the $\pm 2\sigma$ error bars on the measured power spectrum.
Where the boxes meet the horizontal axis we are consistent with zero.
The thin red line is the 1-$\sigma$ noise level, and 
the magenta line is the 2-$\sigma$ upper limit for each $k$ bin. 
A fiducial theoretical model for a fully neutral IGM from \citet{Furlanetto:2006} is shown
in black for reference.
\label{fig:1d_ps}
}
\end{center}
\end{figure*}

The resulting one dimensional power spectra are shown in Figure~\ref{fig:1d_ps}, where 
the measured power is shown with solid blue, the 1-$\sigma$ noise level is shown with 
thin red, and the 2-$\sigma$ upper limit for each bin is shown with magenta. 
Where our unbiased estimator is negative, the absolute value is shown, but with a dotted 
blue line. For consistency with the literature, we plot our one dimensional power spectra 
as $\Delta^2(k)=k^3P_{21}(k)/(2\pi^2)$, which has units of mK$^2$.

In all bands and polarizations, we are heavily signal dominated in our most sensitive region
(low $k$). This is not surprising based on the 2D spectra we examined earlier. The gaps in
the data are due to the excised coarse band harmonics. We can see the cable reflection line
between the first two gaps in all polarizations and sub-bands. Between the coarse band
harmonic lines, especially at larger $k$, are bins which approach the noise level and are
consistent with zero (notably the North-South polarization for the low and mid bands). The
bin just above the first coarse band harmonic is also consistent with zero in the low band.
These regions are encouraging because they have the potential to continue integrating down
with more data without further analysis improvements. Further suppression of the coarse band harmonic and cable reflection lines
will also improve the limits presented here.

Our best upper limits for a cosmological signal are at low $k$, despite the 
strong leakage. Because any foreground leakage should not correlate with the EoR signal, 
we can assume that the power of the sum of leakage and cosmological signal is greater 
than the power of the EoR signal alone, allowing us to set an upper limit. We quote the best 
2-$\sigma$ upper limits, $\Delta^2_{\text{UL}}$, for each polarization and band in Table~
\ref{tbl:limits}. These limits and their context in the field are further discussed in Section~
\ref{sec:discussion}.

\begin{deluxetable}{ccccccc}
\tabletypesize{\footnotesize}
\tablewidth{\columnwidth}
\tablecaption{Upper limits on the EoR power spectrum for our three 
sub-bands and two polarizations. Upper limits, $\Delta^2_{\text{UL}}$, are at 97.7\% 
confidence level. Cosmological wavenumbers, $k$, are in units of h~Mpc$^{-1}$ and upper limits are in units of mK$^2$. The last two columns, RTS$+$CHIPS, are produced from the reference pipeline which is discussed in Section~\ref{subsec:ref_pipe}.
\label{tbl:limits}
}
\tablehead{
& & & \multicolumn{2}{c}{FHD$+$\eppsilon} & \multicolumn{2}{c}{RTS$+$CHIPS} \\
Sub-band & $z_0$ & Pol & $k$  & $\Delta^2_{\text{UL}}$ & $k$ & $\Delta^2_{\text{UL}}$
}
\startdata
Low & 7.1 & E-W & 0.231 & 3.67 $\times 10^4$ & &\\
Low & 7.1 & N-S & 0.27 & 2.70 $\times 10^4$ & 0.16 & 3.2 $\times 10^4$ \\
Mid & 6.8 & E-W & 0.24 & 3.56 $\times 10^4$ & &\\
Mid & 6.8 & N-S & 0.24 & 3.02 $\times 10^4$ & 0.14 & 2.6 $\times 10^4$ \\
High & 6.5 & E-W & 0.20 & 4.70 $\times 10^4$ & &\\
High & 6.5 & N-S & 0.24 & 3.22 $\times 10^4$ & 0.14 & 2.5 $\times 10^4$ \\
\enddata
\end{deluxetable}

\subsection{Comparison with reference pipeline}\label{subsec:ref_pipe}

As demonstrated in \citet{Jacobs:2016}, comparison between independent 
analysis pipelines is essential to evaluate the validity of algorithms under active development. 
Such comparison provides insight into the strengths and weaknesses of the pipelines, as well 
as confirmation of results. The analysis discussed so far has been based on the FHD to 
\eppsilon pipeline, and here we compare with the Real Time System \citep[RTS,][]
{Mitchell:2008,Ord:2010} to Cosmological \ion{H}{1} Power Spectrum \citep[CHIPS,][]
{Trott:2016} pipeline. Below we present a brief overview of the reference pipeline and 
highlight lessons learned from the comparison.

The RTS uses a fundamentally different calibration and foreground subtraction approach than 
FHD, as well as a different input point source model. 
The primary differences lie in the use of ionospheric corrections, the
bandpass calibration, and the lack of a diffuse model. 
RTS accomplishes the interferometric calibration in two steps: a
direction-independent calibration to a single compound calibrator
created by combining the 1000 apparently brightest sources in the sky,
followed by an individual calibration and subtraction of those sources
inside the Calibration Measurement Loop (CML, \citealt{Mitchell:2008}).
The CML estimates individual ionospheric corrections for all sources and
performs a full direction-dependent calibration for the 5 very
brightest ones. These tasks
are distributed over independent parallel threads for each 1.28~MHz
coarse channel.

The bandpass calibration is also fundamentally different to FHD. The bandpass for each antenna is 
fitted to a 3rd order polynomial within each 1.28 MHz coarse band and each direction and 
polarization individually. The delay of the 150~m cable reflection discussed earlier is 
comparable to the scale of the polynomial fit, so we expect it to naturally be fit out of the gain
solutions. Indeed the RTS has not yet seen evidence of the reflection in the processed
data \citep{Jacobs:2016}.

The source catalog used with RTS was a combined cross-match created with the Positional 
Update and Matching Algorithm (PUMA, J.~L.~B.~Line et al. 2016, in review). Full details of PUMA and 
this cross-matched catalog can be found in Line et al. (2016), but we include a brief 
description here. PUMA is designed to cross-match low radio-frequency ($\lesssim 1\,$GHz) 
catalogs. It utilizes a Bayesian probabilistic positional cross-match approach based on~
\citet{Budavari:2008}, combined with criteria based on catalog resolution and fitting the 
spectral energy distribution to a power-law. These criteria allow for surveys with differing 
resolutions, which can introduce confusing sources into the matched process.
Primarily the source catalog was based on the MWA Commissioning Survey
\citep[MWACS, ][]{Hurley-Walker:2014}, which was cross-matched to the $74\,$MHz Very 
Large Array Low Frequency Sky Survey redux~\citep[VLSSr, ][]{Lane:2012}, the Molonglo 
Reference Catalogue~\citep[MRC, ][]{Large:1981}, the $843\,$MHz Sydney University 
Molonglo Sky Survey~\citep[SUMSS,][]{Mauch:2003}; and the $1.4\,$GHz NRAO VLA Sky 
Survey~\citep[NVSS,][]{Condon:1998}. MWACS covers approximately 
$20.5^h < \textrm{RA} < 8.5^h$,  $-ˆ'58^\circ < \delta < -ˆ'14^\circ$, so to complete the sky 
coverage needed, MRC was used as a base catalogue outside of the MWACS coverage and 
was matched to VLSSr, SUMSS and NVSS.

Since RTS has the added step of individual source peeling, there is a
greater possibility for numerical calibration failure. Before passing to CHIPS, a simple quality
assurance step is considered, which is based upon the variance of residual visibilities in each
1.28~MHz coarse band. 
Observations with divergent or
unstable calibrations were readily identified by variances far outside
the otherwise observed distribution of values.
Of the 1,029 observations calibrated and used by the FHD pipeline, the RTS$+$CHIPS 
pipeline successfully calibrated and included 1,003. The remainder failed to calibrate 
successfully, potentially due to adverse ionospheric conditions. 

The CHIPS power 
spectrum estimator applies a full maximum-likelihood estimator to the data. In its full form, the 
data are weighted by the thermal noise, and a model for the residual point source signal, 
within a full frequency-dependent description of the instrument. For this comparison, the 
foreground weighting is not used, in order for a direct comparison with the weighting scheme 
used by \eppsilon.
The data were divided into the 
same three redshift bins, and the 2D and 1D power spectra formed. 

\begin{figure}
\includegraphics[width=\columnwidth]{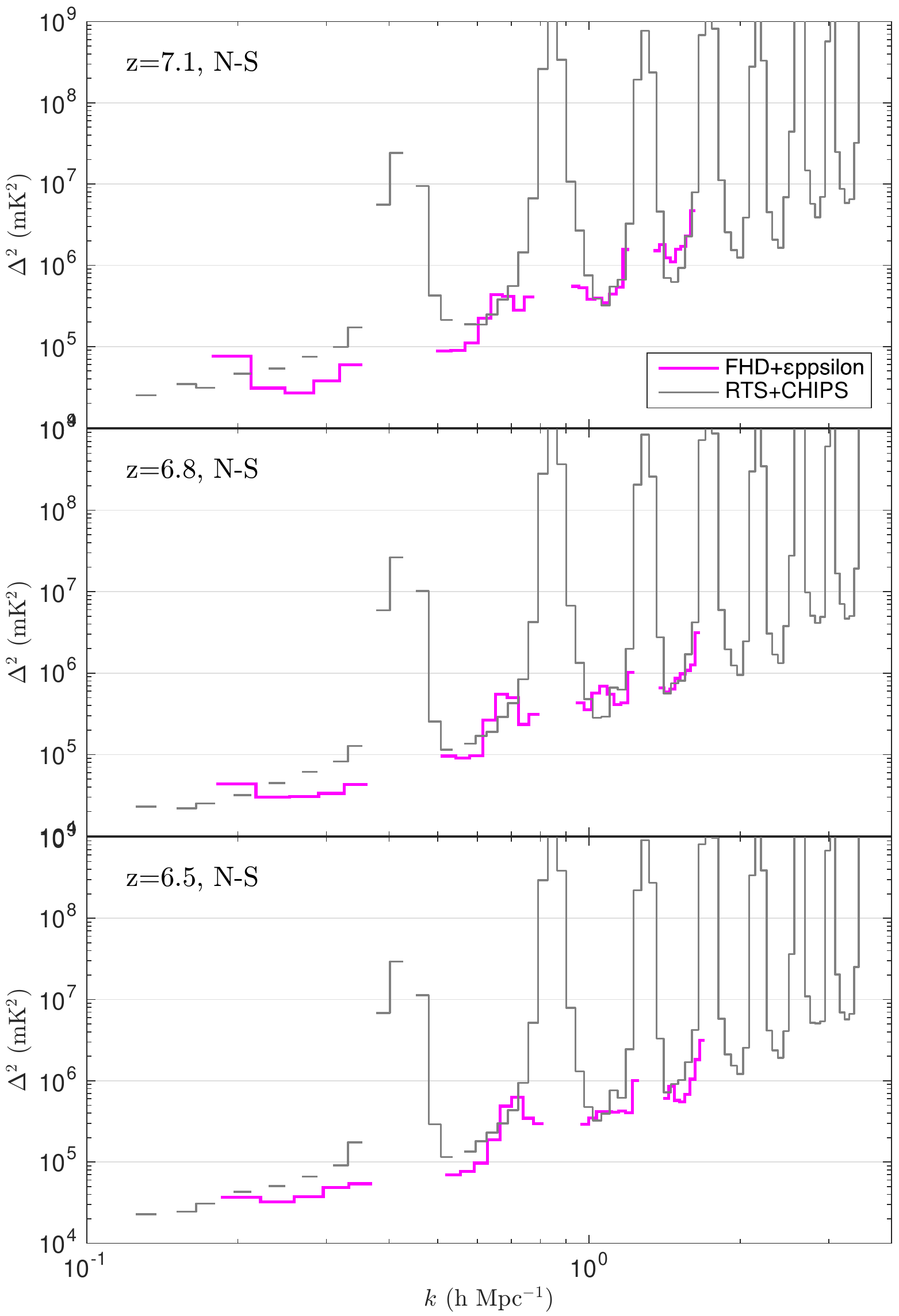}
\caption{
Here we compare the upper limits from the FHD$+$\eppsilon pipeline with those from the 
RTS$+$CHIPS pipeline. The magenta lines are repeated from Figure~\ref{fig:1d_ps}, and the gray lines
are the power plus two sigma thermal error bars lines generated by CHIPS. 
Due to the logarithmic binning used by CHIPS, some bins are empty, resulting in gaps in the
spectra.
CHIPS does not 
excise the coarse band harmonic lines, and so we see their strong effect where there are 
gaps in the magenta lines. 
The RTS$+$CHIPS pipeline accesses larger $k$ modes because the data is
processed at 80~kHz, whereas the FHD$+$\eppsilon pipeline averages to 160~kHz when
gridding.
The most notable differences are the ability of the RTS$+$CHIPS pipe
to access lower $k$ modes, while the FHD$+$\eppsilon pipe suppresses more power before and
after the first coarse band harmonic. While the different analysis methods result in slightly
different upper limits, they are generally in agreement. 
\label{fig:ps_comparison}
}
\end{figure}

Table \ref{tbl:limits} 
displays comparison 2-sigma upper limits, demonstrating broad consistency between the 
pipelines. The RTS$+$CHIPS best limits appear at a lower $k$ bin than FHD$+$\eppsilon, 
owing to the different structure of the foreground leakage for the two pipelines. 
Figure~\ref{fig:slices}, which shows the resulting FHD$+$\eppsilon slices in $k_\parallel$ and 
$k_\bot$, also displays the equivalent power profiles for RTS$+$CHIPS. The comparison 
pipeline shows significantly more power at low $k_\bot$, consistent with no diffuse model 
being subtracted from the data. Aside from this area, the two pipelines show consistent 
results. Figure~\ref{fig:ps_comparison} compares the 1D power from both pipelines for the 
three central redshifts studied in this work. The magenta line reproduces the FHD$+$\eppsilon 
results from Figure~\ref{fig:1d_ps}, while the gray line plots the equivalent RTS$+$CHIPS 
results (with coarse band harmonics included). Despite the differences in the two pipelines, 
they produce consistent power across a range of scales, supporting the robustness of the 
results.

\section{Discussion}\label{sec:discussion}
Inspecting our 32 hour integrated power spectra we see two places where we are limited. 
First there is a region that is leakage-dominated at low $k$. The cause of this leakage is yet 
unknown, but likely due to imperfect calibration, beam models, and/or foreground models.
In order to improve on our limit in this regime, future analysis will require calibration which 
better accounts for the instrumental response, perhaps by increasing signal to noise 
through multiple-snapshot solutions or refined antenna response models. 
Through simulation, \citet{Thyagarajan:2016} recently demonstrated the necessity of
 precise foregrounds and instrumental models in order to retain a clean EoR window.
Studies of the 
beam are underway \citep[e.g.][]{Neben:2015, Sutinjo:2015}, and will likely help to further isolate the 
foreground contaminates in our power spectra. The foreground model is a constant area of 
investigation, but improved spectral dependence in both the point source catalog and the 
diffuse model, as well as the addition of an all-sky Galactic model, will continue to improve 
the foreground subtraction and further unlock the EoR window.

The second limited region in our 1D power spectra is between the coarse band harmonics 
where our measured signal approaches the noise level. Additional data will likely reduce the 
noise, and therefore our upper limit, in this regime. However, until the leakage at low $k$ is 
understood, progress at higher $k$ is susceptible to running into the same systematic
with longer integrations. 

Finally, we place our best upper limits in context with the 21cm EoR field as a whole. A 
direct comparison between measurements from different instruments is difficult due to the 
varying methods, redshifts, and scales probed. However, if we approximate the power 
spectrum in units of mK$^2$ to be roughly flat over the scales probed by the MWA, PAPER, 
and GMRT, we can glean a view of the current state of the field. 
The best result from this analysis is an upper limit on the power spectrum of 
$\Delta^2 \le 2.7 \times 10^4$~mK$^2$ at $k=0.27\, h$~Mpc$^{-1}$ and $z=7.1$.
This is a modest improvement over the previous best MWA results \citep{Dillon:2015}, 
which is not surprising given that this integration contains about ten times more data, but 
is systematics limited. Our results are also consistent with the 
40 hour GMRT limit of $\Delta^2\le6.15\times10^4$~mK$^2$ at $k=0.5\, h$~Mpc$^{-1}$ and $z=8.6$ \citep{Paciga:2013}, though 
probe significantly different periods of the EoR. The frontrunner in terms of an upper limit is 
the PAPER experiment \citep{Ali:2015, Jacobs:2015, Parsons:2014}. The PAPER-64 limit of 
$\Delta^2\le502$~mK$^2$ at $z=8.4$ from a 135 day observing campaign is about one and a half orders of magnitude lower 
than our best limit in power units, though again probing a significantly different redshift. 
The MWA analysis is rapidly improving as we uncover and mitigate systematics to realize
the full sensitivity of the telescope.
The red lines in Figure~\ref{fig:1d_ps} indicate the noise level possible 
to reach with these data if systematics can be overcome.

\subsection{Summary of planned analysis improvements}
Through the reduction of the first season of MWA high band EoR observations, we have
identified several opportunities to improve our analysis. Here we compile a list of future
improvements and their potential impact on the power spectrum results. Broadly speaking,
these improvements emphasize a need to better understand the foregrounds and the 
instrument, and are likely to be relevant to imaging based analyses in general.

\textbf{Extended point source model.} The hierarchical catalog used here was a definite
improvement over the MWACS catalog alone. This point source model will continue
to improve with the release of the GLEAM Survey, as well as dedicated MWA observations 
to target sources in the sidelobes of the EoR fields. Improving the foreground catalog
in our sidelobes will be crucial in controlling the foreground leakage into the EoR window
\citep{Pober:2016}. Incorporating these improved catalogs
into our analysis will improve both the foreground subtraction and the calibration.

\textbf{Diffuse emission model.} The diffuse model here contained no spectral information
and represented only the Stokes I polarization. Nevertheless subtracting this model removed
70\% of the residual power in the foreground wedge. Adding a spectral index and multiple
polarization components to the model will improve this subtraction even further.
As \citet{Lenc:2016} demonstrated, substantial polarized diffuse structure
exists in the EoR0 field. The contributions to Stokes Q and U depend on the ionospheric
conditions, which will be accounted for in future analyses to appropriately model this
emission. 
In addition the diffuse model will be extended to an all-sky model. It is evident from the
jackknives shown in Figure~\ref{fig:jackknife} that strong emission near the horizon is
leaking into our power spectra during off-zenith observations. By modeling and subtracting
this emission we may be able to recover these observations in future analyses.

\textbf{Primary beam model.} In order to properly model the instrument response to the sky
we also need an accurate primary beam model. This is especially important for EoR
power spectrum measurements because the (often bright) near-horizon emission resides
at the edge of the foreground wedge 
\citep{Pober:2016, Thyagarajan:2015, Thyagarajan:2015b}, but the low elevation beam
can be difficult to model at the necessary precision \citep{Thyagarajan:2016}. 
\citet{Asad:2016} demonstrated the need for a polarized primary beam model in order
to mitigate the risk of diffuse polarized emission leaking into Stokes I, conflating with a 
potential cosmological measurement. The MWA
primary beam model is continuously improving to incorporate mutual coupling between the 
phased dipoles, and embedded element patterns for each of the 32 dipoles on each antenna 
\citep{Sutinjo:2015}. Other methods are also being pursued to directly measure the full beam 
response in situ using the ORBCOMM satellite constellation \citep{Neben:2015}, and 
drone-flown calibrator sources (D.~C.~Jacobs, et al. in prep). The latter method is similar to
the strategy \citet{Virone:2014} used to verify SKA beam patterns.

\textbf{Calibration.} The most limiting factor in our analysis is likely to be insufficiently
precise calibration solutions, which can cause foreground power to leak into the EoR window,
as we see in the integrated 2D power spectra (Figure~\ref{fig:2d_ps_subband}).
Because our calibration is based on sky and instrument models, the improvements
described above will improve our calibration. We also plan to combine many snapshot
observations when estimating gains, achieving higher signal to noise, which in turn will
enable more sophisticated parameterizations of the antenna bandpasses. This will require
an overhaul of the data flow depicted in Figure~\ref{fig:pipe} because snapshots will need
to be combined in the calibration step, but preliminary investigations have shown that the
gains are stable enough to be fit across several hours of observation.

Between 2013 and the time of writing, the MWA has collected over two thousand 
hours of data targeting the EoR fields. These observations include the three designated sky 
fields and two frequency bands. In addition, observations are underway to further 
characterize the sources in the instrument sidelobes. The second phase of the MWA
will be commissioned in the latter half of 2016, and will include two sets of highly redundant
cores of antennas. This hybrid configuration will enable a unique opportunity to study
the calibration techniques used in imaging and delay spectrum analyses, providing
another view of the systematics limiting our current results.

While the analysis presented here has not reached full potential, it represents the deepest power spectrum 
integration to date produced by an imaging pipeline. Imaging involves many difficulties (e.g. 
efficient gridding and mapmaking, foreground modeling), but if systematics can be 
overcome, it has the potential to compete with other more targeted experiments and 
analysis styles. 
More broadly, imaging analyses will be necessary to perform cross-correlation studies
with complimentary probes such as galaxy surveys or other intensity mapping experiments
 \citep[e.g.][]{DeBoer:2016, Vrbanec:2016, Beardsley:2015, Silva:2015, Dore:2014, Lidz:2009}, which will ultimately unlock the full potential of 21cm observations.
Here we have identified a crucial region of power spectrum space that is 
currently contaminated, with suggestions for improvement in future analysis. With improved 
calibration techniques, primary beam models, and understanding of foregrounds, the MWA 
and other imaging analyses will be able to quickly approach the most competitive upper 
limits in the field.

\section{Acknowledgements}
This work was supported by NSF grants AST-1410484 and AST-1206552.
DCJ is supported by an NSF Astronomy and Astrophysics Postdoctoral Fellowship
under award AST-1401708.
This scientific work makes use of the Murchison Radio-astronomy Observatory, operated by 
CSIRO. We acknowledge the Wajarri Yamatji people as the traditional owners of the 
Observatory site. Support for the operation of the MWA is provided by the Australian 
Government (NCRIS), under a contract to Curtin University administered by Astronomy 
Australia Limited. We acknowledge the Pawsey Supercomputing Centre which is supported 
by the Western Australian and Australian Governments.

\bibliography{bib}

\end{document}